\documentclass[manuscript]{aastex61}

\shorttitle{Investigation of the $P-M$ Relations for W UMa Systems}
\usepackage{lineno}
\usepackage{amsmath}
\usepackage{graphicx}
\usepackage{xcolor}
\usepackage{longtable}
\usepackage{times}
\usepackage{tablefootnote}
\usepackage{appendix}
\usepackage{lineno}

\begin{document}

\title{Investigation of the Orbital Period and Mass Relations for W UMa-type Contact Systems}

\correspondingauthor{Atila Poro}
\email{astronomydep@raderonlab.ca}

\author{A. Poro}
\affil{Astronomy Department of the Raderon Lab., Burnaby, BC, Canada}

\author{S. Sarabi}
\affil{Astronomy Department of the Raderon Lab., Burnaby, BC, Canada}

\author{S. Zamanpour}
\affil{Binary Systems of South and North (BSN-Project), Educational-Research Department 2021, Iran}

\author{S. Fotouhi}
\affil{Binary Systems of South and North (BSN-Project), Educational-Research Department 2021, Iran}

\author{F. Davoudi}
\affil{Astronomy Department of the Raderon Lab., Burnaby, BC, Canada}

\author{S. Khakpash}
\affil{Department of Physics and Astronomy, University of Delaware, Newark, DE 19716, USA}

\author{S. Ranjbar Salehian}
\affil{Binary Systems of South and North (BSN-Project), Educational-Research Department 2021, Iran}

\author{T. Madayen}
\affil{Astronomy and Astrophysics Department, University of Toronto, Toronto, Canada}

\author{A. Foroutanfar}
\affil{Binary Systems of South and North (BSN-Project), Educational-Research Department 2021, Iran}

\author{E. Bakhshi}
\affil{Department of Physics, University of Zanjan, Zanjan, Iran}

\author{N.S. Mahdavi}
\affil{Bkaran Observatory of Kerman, Kerman, Iran}

\author{F. Alicavus}
\affil{Çanakkale Onsekiz Mart University, Faculty of Arts and Sciences, Department of Physics, 17020, Çanakkale, Turkey}
\affil{Çanakkale Onsekiz Mart University, Astrophysics Research Center and Ulupınar Observatory, 17020, Çanakkale, Turkey}

\author{A. Mazidabadi Farahani}
\affil{Binary Systems of South and North (BSN-Project), Educational-Research Department 2021, Iran}

\author{G. Sabbaghian}
\affil{Department of Chemical Engineering, Science and Research Branch, Islamic Azad University, Tehran, Iran}

\author{R.S. Hosseini}
\affil{Department of Geology, Shahid Bahonar University of Kerman, Kerman, Iran}

\author{A. Aryaeefar}
\affil{Department of Energy Engineering and Physics, Amirkabir University of Technology, Tehran, Iran}

\author{M. Hemati}
\affil{Faculty of Physics, Shahid Bahonar University of Kerman, Kerman, Iran}

\date{Accepted 2021 December 23. Received 2021 December 13; in original form 2021 October 12}

\begin{abstract}

New relationships between the orbital period and some parameters of W Ursae Majoris (W UMa) type systems are presented in this study. To investigate the relationships, we calculated the absolute parameters of a sample of 118 systems. For this purpose, we used the parallax values obtained from the Gaia Early Data Release 3 (Gaia EDR3) star catalog for more precise calculations. The other required parameters, including the light curve solutions and the orbital period were derived from previous research. For some relationships, we added 86 systems from another study with an orbital period of less than 0.6 days to our sample, allowing us to increase the number of systems to 204. Therefore, the mass ($M$) values of each component along with all the other absolute parameters were recalculated for these contact systems. We used the Markov Chain Monte Carlo (MCMC) approach in order to gain the new orbital period-mass relations ($P-M$) per component, and added the temperature ($T$) to the process to acquire the new orbital period-temperature ($P-T_{\rm1}$) relation. We presented the orbital period behavior in terms of $log(g)$ by new relations for each component. We have also obtained a model between the orbital period, the mass of the primary component and temperature ($P-M_{\rm1}-T_{\rm1}$) using the Artificial Neural Networks (ANN) method. Additionally, we present a model for the relationship between the orbital period and the mass ratio ($P-q$) by fitting a Multi-Layer Perceptron (MLP) regression model to a sample of the data collected from the literature.
\end{abstract}


\section{Introduction}\label{sect1}
W UMa-type systems are important astrophysical tools for studying the star formation, stellar structure, and evolution. Therefore, studying them provides information about fundamental stellar parameters and stellar evolution such as mass, temperature, and surface gravity.

The orbital period of contact systems can be related to the absolute parameters (\citealt{2003MNRAS.342.1260Q}). To understand general evolutionary features in comparison with theoretical models, numerous studies have been performed to analyze the orbital period versus the absolute parameters statistically.
Previous studies have shown that there is an orbital period-mass relationship in the contact binary systems. \cite{2003MNRAS.342.1260Q} presented two $P-M_{1}$ relations for A-type and W-type systems. In the study of mass distribution performed by \cite{2006MNRAS.370L..29G}, the results showed that the mass of the primary components increases steeply with an increasing orbital period. In contrast, the mass of secondary components is approximately orbital period independent. \cite{2006MNRAS.373.1483E} estimated the logarithmic format of the $P-M_{\rm tot}$ diagram.
The study of the orbital period by \cite{2008MNRAS.390.1577G} showed that the knowledge of the orbital period alone is adequate to determine the masses of the components. According to \cite{2015AJ....150...69Y}, the relationship between $P-M_{\rm tot}$ implies that mass may escape from the central system when the orbital period decreases. Based on the results of an orbital period-mass relation, the long-period binaries differ from those with a short-period in terms of distribution (\citealt{2018PASJ...70...90K}). \cite{2021ApJS..254...10L} calculated two relations between orbital period and mass for both primary and secondary components. 
\cite{2017RAA....17...87Q} discussed that the correlation between the orbital period and $log(g)$ for short-period EWs is weak and the relationship for these systems is deeper than for long-period ones.

Although these studies have broadened our view of W UMa characteristics, there exists a lack of sufficient focus on some absolute parameters in triple or more relationships. On the other hand, the evolutionary states of EWs are mostly dependent on their orbital periods (\citealt{2017RAA....17...87Q}).

It is also interesting to use statistical methods to infer relations between parameters of binaries without having to fully analyze the systems. This
has been done by studying samples of binaries to find a relationship between their orbital period and mass ratio using different models. \cite{2020MNRAS.497.3493Z}
presented formulations to describe how the orbital period is related to the semi-major axis ($a$) and the mass ratio by using the simplified Kepler's third law and fitting models to the data of a sample of contact binaries.
In this study, the primary goal is to recalculate the absolute parameters of a sample of contact systems based on the Gaia EDR3 parallax. As the next goal, the relationships between $P-M$, $P-T_{\rm1}$, and $P-log(g)$ are updated and investigated using the MCMC approach. Additionally, more innovative methods like the ANN model and the MLP regression model are used to explore relations for $P-M_{\rm1}-T_{\rm1}$ and $P-q$.

\section{Dataset}\label{sec2}
W UMa-type systems have a short orbital period, less than one day (\citealt{2005MNRAS.357..497B}). Several orbital period ranges for contact binaries have been proposed in various studies, each of which with specific reasons. \cite{2003MNRAS.342.1260Q} showed that most of the A-type systems have periods longer than 0.41 days, while the W-type systems have shorter periods. Later, \cite{2017RAA....17...87Q} evaluated a larger sample and discussed that the range of $0.2<P<0.6$ days contains a large number of contact binaries. Another study by \cite{2017RAA....17..115J} declared that the period of most of the contact systems having solar-like components falls in the range of $0.2<P<0.7$ days. \cite{2018PASJ...70...90K} proposed $P<0.6$ days for the majority of contact binaries based on the mass-orbital period diagram of the sample. Also, \cite{2018ApJ...859..140C} studied short period binaries and excluded early-type systems with periods longer than P$\thicksim$0.56 days. The investigation by \cite{2020MNRAS.493.4045J} revealed that the distribution of early-type systems is mostly in the range of  $P>0.5$ days and that their number is less than late-type systems whose period is predominantly shorter than 0.5 days. Eventually, \cite{2021ApJS..254...10L} reported that systems with $P>0.5$ days and $T>7000$ K probably have radiative envelopes and should not be designated as W UMa-type binaries.
We studied 118 W UMa systems, A-type or W-type in the northern and southern celestial hemispheres. Despite different period cuts between the early-type and late-type systems reported in some of the previous studies, we adopted a range of $P<0.6$ days for our sample of W UMa systems. The average star temperature of our dataset is 6057 K for the primary stars and 5848 K for both components.

The apparent magnitude of these systems in the $V$ filter varies between 7$^m$ and 17$^m$. $V_{max}$ values were obtained from the AAVSO Photometric All Sky Survey DR9 (APASS9) catalog (\citealt{2018A&A...616A...1G}) and the All Sky Automated Survey (ASAS) catalog of Variable Stars (\citealt{2002AcA....52..397P}). The Simbad\footnote{http://simbad.u-strasbg.fr/simbad/}  database was used for a small number of systems that were not included in the APASS9 and ASAS catalogs. All the 118 systems have light curve solutions in the previous studies, which were done between 2003 and 2021. We calculated the absolute parameters using some of the light curve solutions parameters in previous studies, together with the Gaia EDR3 parallax; so the validity of the selected studies was important.

We used other samples in this investigation and added them to our sample to achieve reliable results. For this reason, we used a sample of 86 systems of the \cite{2018PASJ...70...90K} study for some relationships, such as $P-M_{\rm1}$, $P-T_{\rm1}$, and $P-M_{\rm1}-T_{\rm1}$ and increased the total number of used samples to 204. Therefore, for other relationships, the \cite{2018PASJ...70...90K} study did not have directly the parameters needed to be added to our sample, and only our 118 systems were used for $P-M_{\rm2}$, $P-M_{\rm tot}$, and $P-log(g)_{\rm1,2}$ relationships. Additionally, we employed the \cite{2020MNRAS.497.3493Z} study sample, as well as our 118 systems, which summed up a total of 369 systems, to investigate $P-q$  relationship.

\section{Method}\label{sec3}
The distance of our systems was calculated from the parallax value of the Gaia EDR3\footnote{https://gea.esac.esa.int/archive/} catalog. When only photometric data is available, the Gaia parallaxes can be used to obtain better estimations of the absolute parameters such as mass of the components, 
absolute and bolometric magnitude, luminosity, and radius all based on observational and light curve solutions. So, the photometric solutions, including $l/l_{\rm tot}$, $r_{\rm (mean)}$, $T$, $P$ and $q$ for each component were collected from valid previous published studies (Table \ref{tab9}).

The Gaia space telescope was launched by the European Space Agency (ESA) in 2013. A systematic update was required in 2018 due to the low accuracy in measuring the stellar parameters. Although the values of positions, parallax, proper motion, temperature, and luminosity of over 1.3 billion stars were the result of this update, it came along with some limitations. These included an increase in observational noise, a few spurious sources, lack of high accuracy of the parallax value, and incapability to measure high proper motion objects precisely (\citealt{2020PASP..132g4501R}). Hence, the system was updated to Gaia EDR3 in 2020 leading to a more complete Celestial Reference Frame (CRF), increasing the accuracy of parallax values by 30\%, and more reliable celestial
positions (\citealt{2021A&A...649A...1G}). An accurate parallax value for a star results in a trustworthy distance (\citealt{2009ApJ...705.1548R}). 
Consequently, this distance brings about a precise measurement of the absolute stellar parameters. Thus, the highly reliable parallax value reached in Gaia EDR3 inspired us to apply it in this study. However, some recent researches have investigated the systematic zero-point offset of the Gaia EDR3 parallaxes, which yielded that the parallax values require correction in order to be less biased and more accurate. \cite{2021A&A...649A...4L} revealed that offering a general recipe for the correction of parallaxes is impossible. They have found that the parallax bias is at least dependent to the other parameters such as magnitude, color, and ecliptic latitude. The corrected parallaxes calculated by \cite{2021A&A...649A...4L} takes quasars into account for fainter sources, then expands the results by differential methods using binaries and the Large Magellanic Cloud objects. \cite{2021ApJ...911L..20R} reached better offsets from the Gaia team official corrected parallaxes and generalized the results to the Galactic objects by deploying W Ursae Majoris (EW)-type eclipsing binary systems. They detected the bias to be less than 10$\mu$as for the corrected parallax values across 40\% of the sky. Although \cite{2021ApJ...911L..20R} confirmed that the parallaxes after correction have smaller bias, their method implies the restriction to magnitude, color, extinction, and a special period interval, that does not cover all the star systems of our sample. As the 118 systems in our dataset are from a large variety of the parameters listed above, we decided to use the Gaia parallaxes before correction. We assured this by checking the correction for the systems, and observed that the results derived from the corrected parallaxes do not exceed the uncertainty range of the mass value.

We investigated relations including $P-M_{\rm1,2}$, $P-M_{\rm tot}$, $P-T_{\rm1}$, and $P-log(g)_{\rm1,2}$ through adding a number of binary systems to graphs. The fitting of the linear and quadratic models to our data was done by the MCMC method to find the parameters of the models and their uncertainties. The lower Bayesian Information Criteria (BIC) favors the linear model over the quadratic model. Thus, we described the relationship between all the quantities by a linear function similar to previous studies.

To find the $P-q$ relationship, we explored different models to investigate their capabilities. We tested a Random Forest Regressor, a Gradient Boosting Regressor, and a Bayesian Ridge Regressor along with a MLP Regressor. Looking at different criteria such as mean squared error and the ranges of the 
predictions, we decided to use the MLP Regressor as the main method for this part as it showed higher levels of flexibility.
A three-parameter $P-M_{\rm1}-T_{\rm1}$ relationship was investigated, and various approaches were considered in artificial intelligence techniques. We concluded the use of the ANN model appears to have the highest degree of compatibility with the sample.

\section{Analysis}\label{sec4}
\subsection{Calculation of the absolute parameters}
The absolute parameters of the binary systems including $M_{\rm v}$, $M_{\rm bol}$, $L$, $R$, and $M$ were calculated. Hence, the necessary parameters such as $q$, $P$, $T$, $r_{\rm (mean)}$, and $l/l_{\rm tot}$ were extracted from the light curve solution of each system. Relation 1 yields the absolute magnitude ($M_{\rm v}$) of each system.

\begin{equation}\label{eq1}
M_v=V-5log(d)+5-A_v
\end{equation}

In relation 1, the extinction coefficient $A_{\rm v}$ and its uncertainty were calculated utilizing the dust-maps Python package of \cite{2019ApJ...887...93G}. 
The absolute magnitude for the primary and the secondary components, $M_{\rm v1}$ and $M_{\rm v2}$, is determined by the following relation.

\begin{equation}\label{eq2}
M_{v(1,2)}-M_{v(tot)}=-2.5log(\frac{l_{(1,2)}}{l_{(tot)}})
\end{equation}

The bolometric magnitude ($M_{\rm bol}$) of each component of the systems is the result of relation 3, where the bolometric correction ($BC$) is taken from \cite{2020MNRAS.496.3887E}.

\begin{equation}\label{eq3}
M_{bol}=M_{v}+BC
\end{equation}

As shown below, the luminosity of the individual components ($L$) is derived from Pogson’s relation (\citealt{1856MNRAS..17...12P}).
$M_{\rm bol\odot}$ is taken as 4.73 mag
(\citealt{2010AJ....140.1158T}).

\begin{equation}\label{eq4}
M_{bol}-M_{bol_{\odot}}=-2.5log(\frac{L}{L_{\odot}})
\end{equation}

Consequently, the radii of the primary and the secondary components $R_{\rm1}$ and $R_{\rm2}$ were calculated by inserting the temperature of each component as $T$ and the Stefan-Boltzmann constant as $\sigma$ in relation 5. The mean fractional radii ($r_{\rm mean}$) of the components were obtained by relation 6, where $r_{\rm back}$, $r_{\rm side}$, and $r_{\rm pole}$ are specified separately in the light curve solution of the systems.

\begin{equation}\label{eq5}
R=(\frac{L}{4\pi \sigma T^{4}})^{1/2}
\end{equation}

\begin{equation}\label{eq6}
r_{mean}=(r_{back} \times r_{side} \times r_{pole})^{1/3}
\end{equation}

According to Kepler’s third law (\citealt{2021AJ....161..221P}), the total mass of the system ($M_{\rm1}+M_{\rm2}$), in terms of $M_{\odot}$, is acquired from
relation 8. The separation $a$ is the average value of $a_{\rm1}$ and $a_{\rm2}$ calculated for each component individually by their radius ($R$) and the mean fractional radius ($r$) using relation 7. However, a multiplication by $R_{\odot}$ must be considered before applying the value of a in relation 8; $G$ is the gravitational constant and $P$ is the orbital period of the system in seconds.

\begin{equation}\label{eq7}
a=\frac{R}{r}
\end{equation}

\begin{equation}\label{eq8}
\frac{a^3}{G(M_1+M_2)}=\frac{P^2}{4\pi^2}
\end{equation}

Finally, the mass summation ($M_{\rm1}+M_{\rm2}$) was calculated, and the mass ratio $q=M_{\rm2}⁄M_{\rm1}$, represented in the light curve solution of each system results in the determination of the mass of each component $M_{\rm1}$ and $M_{\rm2}$.
The absolute parameters calculated in this study are presented in Table \ref{tab}. The uncertainties of the parameters were calculated considering the error bars of the light curve solutions of the systems (e.g. $P$, $l/l_{\rm tot}$, $q$, $T$). If a parameter in the table of solutions did not have an error bar, we have provided the average error bar for that parameter of all systems.

\begin{table*}
\caption{Estimated absolute elements of 118 W UMa systems.} 
\centering
\begin{center}
\footnotesize
\begin{tabular}{c c c c c c c c c c }
 \hline
 \hline
         System&$M_{bol1}^{\ \ mag.}$& $M_{bol2}^{\ \ mag.}$ &$L_{1}(L_{\odot})$ &$L_{2}(L_{\odot})$ &
$R_{1}(R_{\odot})$&$R_{2}(R_{\odot})$ &$a(R_{\odot})$&$M_{1}(M_{\odot})$ &$M_{2}(M_{\odot})$\\
		\hline
        1&
3.824(118)&
5.135(115)&
2.325(243)&	
0.695(61)& 
1.462(154)&
0.805(38)&
2.745(3)&
1.629(53)&
0.335(12) \\
2&
5.587(98)&
5.465(97)&
0.458(27)&
0.516(31)&
0.769(23)&
0.884(33)&
2.133(8)&
0.579(64)&
0.846(90) \\
3&
2.925(59)&
5.089(58)&
5.321(230)&
0.725(26)&
1.703(86)&
0.743(8)&
2.705(7)&
1.464(104)&
0.163(16) \\
4&
5.000(100)&
6.031(266)&
0.787(61)&
0.305(45)&
1.095(54)&
0.752(53)&
2.478(2)&
1.408(33)&
0.503(13) \\
5&
2.992(25)&
4.848(13)&
5.003(105)&
0.905(9)&
1.872(52)&
0.850(9)&
3.261(7)&
2.261(145)&
0.331(25) \\
6&
4.533(88)&
4.022(123)&
1.210(88)&
1.937(200)&
1.032(51)&
1.326(124)&
3.036(6)&
0.609(35)&
1.105(69) \\
7&
5.672(152)&
4.987(205)&
0.424(42)&
0.797(121)&
0.726(36)&
1.189(140)&
2.655(6)&
0.468(36)&
1.985(132) \\
8&
6.616(459)&
6.552(25)&
0.178(49)&
0.188(2)&
0.695(92)&
0.929(2)&
1.7338(9)&
0.570(90)&
0.456(78) \\
9&
4.865(28)&
5.493(26)&
0.891(20)&
0.500(9)&
1.067(17)&
0.942(11)&
2.694(7)&
1.928(149)&
1.157(98) \\
10&
5.446(117)&
5.418(124)&
0.522(42)&
0.536(45)&
0.811(36)&
0.897(44)&
2.251(4)&
0.759(39)&
0.987(55) \\
11&
4.338(144)&
5.136(188)&
1.448(140)&
0.694(74)&
1.254(105)&
0.866(52)&
2.620(12)&
1.217(150)&
0.452(70) \\
12&
3.351(78)&
4.579(87)&
3.594(252)&
1.160(84)&
1.620(126)&
0.941(42)&
3.214(28)&
2.010(454)&
0.615(93) \\
13&
5.161(13)&
4.646(34)&
0.679(7)&
1.090(30)&
0.912(6)&
1.380(35)&
3.151(3)&
0.699(38)&
2.238(166) \\
14&
5.855(86)&
6.309(91)&
0.358(18)&
0.236(11)&
0.812(23)&
0.737(11)&
1.992(21)&
0.850(85)&
0.646(84) \\
15&
5.178(151)&
5.606(162)&
0.668(71)&
0.450(46)&
0.814(49)&
0.771(33)&
2.053(17)&
0.830(64)&
0.690(96) \\
16&
6.540(311)&
6.903(310)&
0.190(25)&
0.136(15)&
0.692(44)&
0.553(24)&
1.718(61)&
0.744(47)&
0.357(59) \\
17&
6.417(96)&
6.226(121)&
0.213(11)&
0.254(17)&
0.710(17)&
0.953(39)&
2.164(9)&
1.642(107)&
0.630(88) \\
18&
5.215(121)&
5.148(132)&
0.646(49)&
0.687(57)&
0.845(37)&
0.781(26)&
2.023(14)&
0.696(48)&
0.577(53) \\
19&
5.663(118)&
5.013(153)&
0.427(33)&
0.778(92)&
0.828(37)&
1.220(117)&
2.929(3)&
0.703(40)&
2.357(75) \\
20&
6.179(176)&
5.254(174)&
0.266(29)&
0.623(75)&
0.513(20)&
0.826(55)&
1.726(1)&
0.183(13)&
0.689(32) \\
21&
7.488(123)&
7.164(122)&
0.080(3)&
0.107(6)&
0.652(12)&
0.774(23)&
1.872(22)&
0.627(73)&
1.351(255) \\
22&
2.364(67)&
4.507(70)&
8.920(440)&
1.240(59)&
2.250(171)&
1.027(31)&
4.011(2)&
2.053(130)&
0.308(15) \\
23&
3.666(755)&
5.221(728)&
2.689(709)&
0.642(513)&
1.570(519)&
0.784(340)&
2.831(3)&
1.677(53)&
0.265(12) \\
24&
3.779(76)&
5.252(97)&
2.423(143)&
0.624(39)&
1.673(113)&
0.931(36)&
3.485(5)&
2.314(199)&
0.317(16) \\
25&
4.119(1)&
6.137(2)&
1.771(2)&
0.276(2)&
1.119(1)&
0.462(1)&
1.897(3)&
0.678(32)&
0.083(10) \\
        \hline
        \hline
\end{tabular}
\end{center}
\label{tab}
\end{table*}
 
\begin{table*}
\renewcommand\thetable{1}
\caption{Continued} 
\centering
\begin{center}
\footnotesize
\begin{tabular}{c c c c c c c c c c }
 \hline
 \hline
         System&$M_{bol1}^{\ \ mag.}$& $M_{bol2}^{\ \ mag.}$ &$L_{1}(L_{\odot})$ &$L_{2}(L_{\odot})$ &
$R_{1}(R_{\odot})$&$R_{2}(R_{\odot})$ &$a(R_{\odot})$&$M_{1}(M_{\odot})$ &$M_{2}(M_{\odot})$\\
		\hline
26&
4.306(77)&
6.216(72)&
1.490(95)&
0.257(16)&
1.171(60)&
0.468(8)&
1.957(1)&
0.842(33)&
0.094(10) \\
27&
5.000(41)&
4.246(204)&
0.787(25)&
1.576(268)&
0.796(14)&
1.169(142)&
2.588(1)&
0.319(16)&
0.854(37) \\
28&
4.045(242)&
4.924(247)&
1.896(374)&
0.844(174)&
1.122(171)&
0.723(72)&
2.327(2)&
0.873(22)&
0.330(10) \\
29&
3.812(164)&
4.715(162)&
2.350(315)&
1.023(139)&
1.318(41)&
0.833(64)&
2.700(2)&
1.176(65)&
0.422(13) \\
30&
5.862(125)&
5.802(114)&
0.356(30)&
0.376(26)&
0.602(8)&
0.763(27)&
1.803(1)&
0.583(19)&
0.619(11) \\
31&
2.508(26)&
4.442(11)&
7.813(136)&
1.316(9)&
1.904(45)&
0.800(4)&
3.277(42)&
2.155(312)&
0.280(49) \\
32&
5.611(66)&
6.177(53)&
0.448(21)&
0.266(11)&
0.958(31)&
0.779(18)&
2.389(1)&
1.069(33)&
0.826(31) \\
33&
3.728(69)&
4.606(81)&
2.540(128)&
1.131(62)&
0.998(34)&
0.685(17)&
2.015(2)&
0.462(14)&
0.174(10) \\
34&
6.340(80)&
6.13(9)&
0.229(11)&
0.278(2)&
0.576(12)&
0.776(3)&
1.986(9)&
0.350(45)&
1.165(170) \\
35&
5.793(43)&
5.209(127)&
0.379(11)&
0.649(65)&
0.571(4)&
0.866(52)&
1.942(10)&
0.296(43)&
1.033(173) \\
36&
4.003(8)&
6.779(9)&
1.971(15)&
0.153(1)&
1.181(8)&
0.294(1)&
1.893(4)&
0.882(56)&
0.058(10) \\
37&
5.645(19)&
6.099(20)&
0.434(5)&
0.286(3)&
0.572(2)&
0.501(3)&
1.492(67)&
0.400(60)&
0.260(37) \\
38&
6.810(21)&
7.062(21)&
0.149(1)&
0.118(1)&
0.680(3)&
0.497(3)&
1.639(53)&
0.835(47)&
0.334(94) \\
39&
5.106(27)&
5.826(25)&
0.714(12)&
0.368(5)&
1.339(20)&
0.965(2)&
3.222(127)&
2.455(368)&
0.982(91) \\
40&
6.292(44)&
5.240(28)&
0.239(5)&
0.631(15)&
0.645(6)&
1.154(27)&
1.897(1)&
0.793(23)&
0.441(17) \\
41&
5.430(48)&
5.937(49)&
0.530(17)&
0.332(12)&
0.824(15)&
0.593(9)&
1.860(1)&
0.693(31)&
0.366(16) \\
42&
5.194(225)&
6.139(223)&
0.658(99)&
0.276(45)&
0.858(76)&
0.520(30)&
1.784(2)&
0.665(22)&
0.235(11) \\
43&
5.618(142)&
6.316(140)&
0.446(37)&
0.234(20)&
0.770(33)&
0.515(16)&
1.678(1)&
0.545(19)&
0.235(10) \\
44&
5.535(30)&
6.271(32)&
0.481(10)&
0.244(6)&
0.794(9)&
0.519(4)&
1.715(2)&
0.534(23)&
0.278(10) \\
45&
5.362(158)&
6.446(157)&
0.564(61)&
0.208(25)&
0.823(50)&
0.464(18)&
1.683(2)&
0.608(21)&
0.188(10) \\
46&
4.213(7)&
5.653(4)&
1.625(9)&
0.431(1)&
1.286(6)&
0.692(3)&
2.465(5)&
1.562(95)&
0.281(20) \\
47&
5.326(73)&
5.621(73)&
0.583(28)&
0.444(19)&
0.761(17)&
0.602(11)&
1.7438(6)&
0.547(56)&
0.296(34) \\
48&
4.132(184)&
4.311(184)&
1.751(268)&
1.485(235)&
1.252(165)&
1.136(131)&
3.194(45)&
1.470(78)&
1.264(148) \\
49&
4.627(56)&
5.666(54)&
1.110(53)&
0.426(20)&
1.048(36)&
0.900(8)&
2.550(18)&
1.229(267)&
1.089(259) \\
50&
4.602(89)&
4.677(91)&
1.135(86)&
1.060(87)&
1.019(54)&
1.062(55)&
2.747(7)&
0.980(73)&
1.179(96) \\
        \hline
        \hline
\end{tabular}
\end{center}
\label{tab}
\end{table*}
 
\begin{table*}
\renewcommand\thetable{1}
\caption{Continued} 
\centering
\begin{center}
\footnotesize
\begin{tabular}{c c c c c c c c c c }
 \hline
 \hline
         System&$M_{bol1}^{\ \ mag.}$& $M_{bol2}^{\ \ mag.}$ &$L_{1}(L_{\odot})$ &$L_{2}(L_{\odot})$ &
$R_{1}(R_{\odot})$&$R_{2}(R_{\odot})$ &$a(R_{\odot})$&$M_{1}(M_{\odot})$ &$M_{2}(M_{\odot})$\\
		\hline
51&
5.072(28)&
4.841(23)&
0.736(17)&
0.911(10)&
0.958(15)&
1.216(12)&
2.541(10)&
1.069(125)&
1.145(146) \\
52&
5.630(43)&
4.730(62)&
0.440(13)&
1.009(55)&
0.649(9)&
1.016(19)&
2.318(4)&
0.434(21)&
1.503(81) \\
53&
3.731(235)&
4.424(233)&
2.533(377)&
1.338(202)&
1.649(279)&
1.180(142)&
3.606(19)&
2.220(358)&
1.063(189) \\
54&
4.853(189)&
5.633(187)&
0.901(116)&
0.439(52)&
0.942(78)&
0.674(36)&
2.058(2)&
0.698(20)&
0.306(10) \\
55&
4.592(248)&
5.883(288)&
1.146(114)&
0.349(57)&
1.116(85)&
0.617(41)&
2.404(20)&
0.984(158)&
0.404(52) \\
56&
6.044(23)&
5.957(112)&
0.301(24)&
0.326(89)&
0.733(5)&
0.814(77)&
1.982(41)&
0.613(70)&
0.856(91) \\
57&
5.106(41)&
4.632(82)&
0.714(22)&
1.105(72)&
1.074(20)&
1.064(35)&
2.794(18)&
1.455(287)&
1.241(268) \\
58&
5.137(275)&
5.120(280)&
0.694(132)&
0.705(73)&
0.912(42)&
0.821(48)&
2.309(2)&
1.185(30)&
0.457(13) \\
59&
6.328(343)&
6.844(345)&
0.232(36)&
0.144(18)&
0.763(62)&
0.609(32)&
1.726(4)&
0.985(69)&
0.473(36) \\
60&
6.064(8)&
5.912(3)&
0.295(19)&
0.340(8)&
0.600(17)&
0.737(34)&
1.748(2)&
0.237(18)&
0.414(15) \\
61&
5.145(17)&
4.884(6)&
0.689(117)&
0.876(54)&
0.721(63)&
0.926(149)&
2.131(4)&
0.485(26)&
0.912(54) \\
62&
5.615(338)&
5.168(356)&
0.447(128)&
0.674(281)&
0.641(216)&
0.877(398)&
1.950(3)&
0.277(19)&
0.638(31) \\
63&
3.160(161)&
3.635(176)&
4.286(562)&
2.767(609)&
1.229(679)&
1.237(603)&
3.237(3)&
1.574(43)&
0.482(20) \\
64&
5.881(97)&
5.920(140)&
0.350(22)&
0.337(30)&
0.632(17)&
0.6915(27)&
1.695(2)&
0.361(17)&
0.482(18) \\
65&
4.049(91)&
4.592(90)&
1.890(142)&
1.146(88)&
1.211(53)&
0.996(41)&
2.571(161)&
0.659(98)&
0.441(65) \\
66&
6.362(141)&
7.341(138)&
0.225(24)&
0.091(8)&
0.602(19)&
0.386(9)&
1.163(11)&
0.264(80)&
0.082(28) \\
67&
6.171(109)&
7.226(107)&
0.268(17)&
0.101(7)&
0.713(12)&
0.456(9)&
1.430(12)&
0.503(89)&
0.211(63) \\
68&
4.195(168)&
5.026(169)&
1.652(197)&
0.768(83)&
1.016(85)&
0.700(36)&
2.159(2)&
0.681(19)&
0.253(11) \\
69&
6.026(758)&
4.349(6)&
0.306(216)&
1.434(8)&
0.465(106)&
1.035(30)&
1.950(1)&
0.098(22)&
0.915(15) \\
70&
4.026(138)&
5.401(137)&
1.930(180)&
0.544(41)&
1.191(91)&
0.633(21)&
2.313(1)&
0.952(25)&
0.180(10) \\
71&
5.873(13)&
6.033(18)&
0.352(3)&
0.304(3)&
0.805(4)&
0.657(4)&
1.763(3)&
0.521(26)&
0.278(15) \\
72&
3.399(129)&
4.801(130)&
3.439(400)&
0.945(91)&
1.400(77)&
0.750(18)&
2.682(19)&
0.998(62)&
0.190(49) \\
73&
4.002(190)&
4.723(192)&
1.973(293)&
1.016(144)&
1.329(77)&
0.963(39)&
2.871(44)&
0.987(71)&
0.448(26) \\
74&
5.291(9)&
4.372(16)&
0.602(4)&
1.404(19)&
0.785(5)&
1.346(32)&
2.535(18)&
0.391(82)&
1.047(147) \\
75&
7.012(52)&
7.637(50)&
0.123(3)&
0.069(1)&
0.795(10)&
0.609(5)&
1.829(19)&
1.364(244)&
0.587(81) \\
        \hline
        \hline
\end{tabular}
\end{center}
\label{tab}
\end{table*}
 
\begin{table*}
\renewcommand\thetable{1}
\caption{Continued} 
\centering
\begin{center}
\footnotesize
\begin{tabular}{c c c c c c c c c c }
 \hline
 \hline
         System&$M_{bol1}^{\ \ mag.}$& $M_{bol2}^{\ \ mag.}$ &$L_{1}(L_{\odot})$ &$L_{2}(L_{\odot})$ &
$R_{1}(R_{\odot})$&$R_{2}(R_{\odot})$ &$a(R_{\odot})$&$M_{1}(M_{\odot})$ &$M_{2}(M_{\odot})$\\
		\hline
76&
3.930(39)&
4.047(53)&
2.109(74)&
1.893(89)&
1.239(37)&
1.215(42)&
3.175(43)&
1.198(103)&
1.073(125) \\
77&
2.797(3)&
3.449(1)&
5.987(14)&
3.284(3)&
1.716(4)&
1.310(41)&
3.898(3)&
2.308(52)&
1.133(68) \\
78&
3.205(59)&
4.597(55)&
4.112(203)&
1.141(48)&
1.822(112)&
1.009(25)&
3.892(4)&
2.854(86)&
0.645(22) \\
79&
6.002(21)&
5.794(21)&
0.313(4)&
0.379(5)&
0.794(6)&
1.004(14)&
2.510(57)&
0.996(77)&
1.634(145) \\
80&
5.784(113)&
5.248(126)&
0.382(28)&
0.626(64)&
0.666(22)&
0.947(63)&
2.258(15)&
0.421(82)&
1.219(267) \\
81&
5.838(175)&
5.484(179)&
0.363(36)&
0.504(65)&
0.710(35)&
0.950(75)&
2.279(1)&
0.449(16)&
1.033(34) \\
82&
6.645(138)&
6.725(165)&
0.173(11)&
0.161(15)&
0.634(17)&
0.654(23)&
1.689(7)&
0.501(61)&
0.775(89) \\
83&
5.170(291)&
4.441(293)&
0.673(138)&
1.317(303)&
0.710(72)&
1.023(165)&
2.220(1)&
0.263(13)&
0.672(19) \\
84&
2.883(55)&
4.899(46)&
5.531(273)&
0.864(31)&
1.562(82)&
1.059(23)&
3.462(11)&
1.649(158)&
0.709(75) \\
85&
4.992(27)&
5.794(23)&
0.790(18)&
0.379(6)&
0.884(14)&
0.575(3)&
1.894(15)&
0.832(107)&
0.254(72) \\
86&
5.420(32)&
5.96(33)&
0.533(11)&
0.324(7)&
0.805(50)&
0.627(32)&
1.805(23)&
0.642(66)&
0.352(63) \\
87&
3.685(72)&
5.678(36)&
2.64(17)&
0.42(1)&
1.64(6)&
0.63(5)&
2.71(12)&
1.546(54)&
0.127(37) \\
88&
5.715(88)&
6.574(85)&
0.407(17)&
0.185(6)&
0.769(17)&
0.516(5)&
1.640(5)&
0.729(67)&
0.213(22) \\
89&
6.445(371)&
6.097(377)&
0.208(33)&
0.287(51)&
0.564(33)&
0.738(67)&
1.704(9)&
0.243(37)&
0.553(95) \\
90&
4.660(117)&
5.244(127)&
1.076(104)&
0.629(59)&
0.999(66)&
0.767(35)&
2.338(13)&
1.609(276)&
0.333(67) \\
91&
2.771(87)&
4.461(87)&
6.132(486)&
1.293(93)&
1.649(148)&
0.828(33)&
3.145(4)&
1.240(47)&
0.238(10) \\
92&
4.609(155)&
4.845(156)&
1.129(128)&
0.908(100)&
1.258(123)&
0.912(61)&
2.840(21)&
1.196(277)&
0.511(31) \\
93&
4.426(39)&
5.577(37)&
1.335(37)&
0.463(10)&
1.119(23)&
0.664(2)&
2.134(1)&
1.067(36)&
0.221(13) \\
94&
2.305(1)&
3.229(3)&
9.419(8)&
4.022(11)&
1.783(3)&
1.172(10)&
4.045(36)&
1.855(224)&
0.668(42) \\
95&
3.499(5)&
4.774(1)&
3.136(12)&
0.969(1)&
1.399(4)&
0.861(2)&
3.073(72)&
1.717(145)&
0.476(48) \\
96&
2.027(11)&
3.653(6)&
12.167(111)&
2.721(14)&
2.377(34)&
1.284(7)&
4.985(26)&
3.727(398)&
0.932(131) \\
97&
4.280(24)&
4.703(23)&
1.528(14)&
1.035(9)&
1.073(7)&
0.829(3)&
2.618(11)&
1.684(213)&
0.899(125) \\
98&
4.118(5)&
5.162(8)&
1.773(8)&
0.678(4)&
1.138(4)&
0.697(4)&
2.521(3)&
1.187(42)&
0.340(13) \\
99&
6.325(158)&
5.532(120)&
0.232(21)&
0.482(42)&
0.513(25)&
0.900(33)&
1.860(2)&
0.161(15)&
0.868(29) \\
100&
6.070(134)&
5.777(109)&
0.294(24)&
0.385(28)&
0.650(24)&
0.837(35)&
1.948(3)&
0.341(16)&
0.712(35) \\
        \hline
        \hline
\end{tabular}
\end{center}
\label{tab}
\end{table*}
 
\begin{table*}
\renewcommand\thetable{1}
\caption{Continued} 
\centering
\begin{center}
\footnotesize
\begin{tabular}{c c c c c c c c c c }
 \hline
 \hline
         System&$M_{bol1}^{\ \ mag.}$& $M_{bol2}^{\ \ mag.}$ &$L_{1}(L_{\odot})$ &$L_{2}(L_{\odot})$ &
$R_{1}(R_{\odot})$&$R_{2}(R_{\odot})$ &$a(R_{\odot})$&$M_{1}(M_{\odot})$ &$M_{2}(M_{\odot})$\\
		\hline
101&
4.539(117)&
5.146(102)&
1.203(112)&
0.688(62)&
1.175(90)&
0.983(61)&
2.865(15)&
1.512(240)&
1.180(205) \\
102&
5.517(153)&
5.414(110)&
0.489(56)&
0.538(45)&
0.759(46)&
0.946(52)&
2.196(4)&
0.455(24)&
0.865(50) \\
103&
4.960(128)&
5.617(100)&
0.817(85)&
0.446(31)&
1.041(78)&
0.832(33)&
2.514(35)&
1.285(290)&
0.668(44) \\
104&
5.069(75)&
4.924(31)&
0.739(43)&
0.844(21)&
0.868(31)&
1.064(20)&
2.592(13)&
0.701(99)&
1.122(184) \\
105&
4.703(89)&
5.090(139)&
1.035(68)&
0.724(68)&
1.415(28)&
1.089(4)&
3.242(2)&
1.882(34)&
0.894(18) \\
106&
4.070(153)&
4.898(153)&
1.853(237)&
0.865(100)&
1.164(119)&
0.763(47)&
2.503(1)&
0.957(31)&
0.251(13) \\
107&
4.663(287)&
3.843(289)&
1.074(225)&
2.2845(519)&
0.886(113)&
1.323(271)&
2.796(1)&
0.486(15)&
1.311(34) \\
108&
3.798(5)&
4.824(5)&
2.381(11)&
0.926(3)&
1.362(7)&
0.835(6)&
2.793(3)&
1.218(39)&
0.362(13) \\
109&
4.095(134)&
4.858(136)&
1.811(205)&
0.897(96)&
1.001(78)&
0.700(28)&
2.182(18)&
0.670(74)&
0.275(80) \\
110&
3.564(16)&
4.072(17)&
2.954(34)&
1.850(21)&
1.401(15)&
1.138(4)&
3.097(6)&
0.989(57)&
0.603(38) \\
111&
3.981(75)&
4.389(76)&
2.012(133)&
1.382(90)&
1.271(73)&
1.089(50)&
3.004(10)&
1.535(152)&
1.064(116) \\
112&
4.152(178)&
5.065(176)&
1.719(236)&
0.741(93)&
1.337(162)&
0.773(52)&
2.644(3)&
0.976(33)&
0.251(10) \\
113&
2.804(45)&
4.379(46)&
5.948(248)&
1.394(55)&
2.018(116)&
1.054(24)&
3.806(26)&
2.697(375)&
0.569(41) \\
114&
6.700(34)&
6.205(37)&
0.164(3)&
0.259(6)&
0.518(3)&
0.688(7)&
1.568(8)&
0.208(31)&
0.542(90) \\
115&
3.989(166)&
5.471(161)&
1.997(272)&
0.510(54)&
1.189(133)&
0.591(25)&
2.175(3)&
0.925(38)&
0.138(11) \\
116&
4.756(5)&
3.980(4)&
0.985(3)&
2.014(4)&
0.876(2)&
1.368(25)&
2.741(1)&
0.709(18)&
1.615(38) \\
117&
4.908(346)&
5.591(347)&
0.857(208)&
0.457(96)&
0.969(157)&
0.654(61)&
2.096(3)&
0.880(37)&
0.301(14) \\
118&
6.319(266)&
5.782(268)&
0.234(31)&
0.383(62)&
0.692(44)&
0.985(105)&
2.166(1)&
0.415(15)&
1.182(37) \\
        \hline
        \hline
\end{tabular}
\end{center}
\label{tab-1}
\end{table*}

\subsection{$P-M$ relationships}
Study of the orbital period variation as a parameter to understand the evolutionary state of contact binaries resulted in finding relations between the orbital period change and parameters such as the mass ratio and the mass of each primary component (\citealt{2001MNRAS.328..635Q}; \citealt{2001MNRAS.328..914Q}). Later studies attempted to continue finding more relations between the orbital period and the absolute parameters.

A study on the $P-M_{\rm1}$ relationship was performed by \cite{2003MNRAS.342.1260Q} on a sample of overcontact binaries (48 W-type and 30 A-type). The results showed that there are two narrower parallel sequences on the $P-M_{\rm1}$ diagram. So, two relations between $P$ and $M_{\rm1}$ were obtained with the least-squares method. The systems with a mass of $M_{\rm1}<1.35 M_{\rm \odot}$ were on the first sequence (both A-type and W-type) and all the systems with a larger mass, $M_{\rm1}>1.35 M_{\rm \odot}$ were on the second sequence (A-type). The significant correlation between
the $P$ and $M_{\rm1}$ diagram revealed that there exists a free matter exchange between the components, based on the Thermal Relaxation Oscillation theory (TRO).

In addition, \cite{2006MNRAS.370L..29G} studied a sample of low-temperature contact binaries (60 A-type, 52 W-type, and 25 near-contact binaries).
Based on the total mass ($M_{\rm tot}$) distribution of the systems, these results show that the mass of the primary components of W UMa-type binaries
increases linearly with increasing orbital period. In contrast, the mass of the secondary component is almost independent of the orbital period and varies between 0 and 1 $M_{\rm \odot}$.

Another study by \cite{2006MNRAS.373.1483E} presented the investigation of 102 W UMa, A-type or W-type systems, estimated the logarithmic format of the $P-M_{\rm tot}$ diagram and calculated an equation accordingly. The results demonstrated that W-type systems dominate over A-types in those with shorter periods.

Then, the \cite{2008MNRAS.390.1577G} analysis of 112 contact binary systems resulted in two $P-M$ relations for both components of the systems. It was
found that determining the orbital period is sufficient for estimating the masses of both components with a reasonable accuracy of about 15\%.

\cite{2015AJ....150...69Y} analyzed a sample of 46 Deep Low Mass Ratio (DLMR) overcontact binaries (i.e., $q\leqslant0.25$ and $f\geqslant50\%$) statistically.
They presented a relationship between $P$ and $M_{\rm tot}$ calculated by the linear least-squares fitting method. The relationship between
$P$ and $M_{\rm tot}$ indicates the decrease in total mass with decreasing orbital period, which implies that the orbit is shrinking and mass is being lost from the central system during its evolutionary process. \cite{2015AJ....150...69Y} mentioned that mass will be lost from the system when the orbital period decreases.

In an investigation, \cite{2018PASJ...70...90K} calculated the primary component masses of 111 overcontact binaries in the Kepler eclipsing binary stars’ catalog based on a mass-temperature relation given by \cite{1988BAICz..39..329H}. By engaging the least-squares method, an orbital period-mass relation for the primary component was calculated for the systems with $P<0.6$ days.

Eventually, an analysis by \cite{2021ApJS..254...10L} on 235 W UMa systems resulted in two relations between the orbital period and mass of both components separately.
Table \ref{tab1} shows the results of the previous studies comprising measured $P-M$ relationships.

\begin{table*}
\caption{Relations between orbital period and mass of components for EW-type systems based on previous studies.}
\centering
\begin{center}
\footnotesize
\begin{tabular}{c c c }
 \hline
 \hline
        Parameters & Relation & Reference \\
	   \hline
$P-M_{\rm 1}$ & $M_{1(W-type)} =0.391(\pm0.059)+1.96(\pm0.17)P$ & \cite{2003MNRAS.342.1260Q} \\ 
$P-M_{\rm 1}$ & $M_{1(A-type)} =0.761(\pm0.150)+1.82(\pm0.28)P$ & \cite{2003MNRAS.342.1260Q} \\ 
$P-M_{\rm 1}$ & $log(M_{1})=(0.755\pm0.059)logP+(0.416\pm0.024)$ & \cite{2008MNRAS.390.1577G} \\
$P-M_{\rm 1}$ & $log(M_{1})=(0.911\pm0.104)logP+(0.475\pm0.044)$ & \cite{2018PASJ...70...90K} \\
$P-M_{\rm 1}$ & $M_{1}=(2.94\pm0.21)P+(0.16\pm0.08)$ & \cite{2021ApJS..254...10L} \\
\hline
$P-M_{\rm2}$ & $log(M_{2})=(0.352\pm0.166)logP-(0.262\pm0.067)$ & \cite{2008MNRAS.390.1577G} \\
$P-M_{\rm2}$ & $M_{2}=(0.15\pm0.17)P+(0.32\pm0.06)$ & \cite{2021ApJS..254...10L}\\
\hline
$P-M_{\rm tot}$ & $log(M_{tot})=0.155logP+0.399$ & \cite{2006MNRAS.373.1483E} \\
$P-M_{\rm tot}$ & $M_{tot}=0.5747(\pm0.0160)+2.3734(\pm0.0331)\times P$ & \cite{2015AJ....150...69Y} \\
       \hline
       \hline
\end{tabular}
\end{center}
\label{tab1}
\end{table*}

According to previous studies on the relationships, in this study we considered all the more massive components as the primary stars.

\begin{figure}
\begin{center}
\includegraphics[width=\columnwidth]{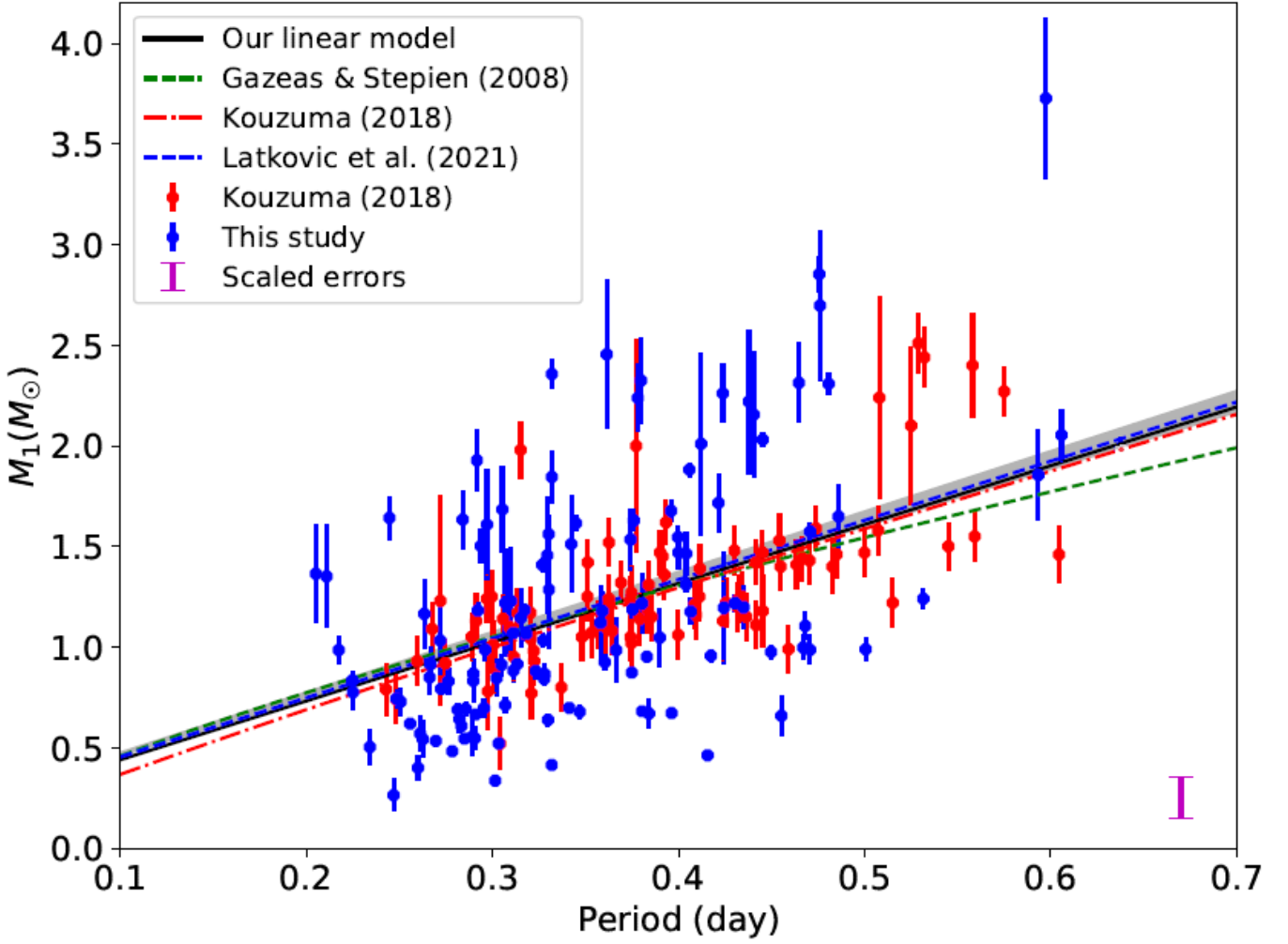}
    \caption{Orbital period versus primary star’s mass. A total of 204 systems from this study and \protect\cite{2018PASJ...70...90K} samples are used in this diagram for our linear model.}
\label{Fig1}
\end{center}
\end{figure}

\begin{figure}
\begin{center}
\includegraphics[width=\columnwidth]{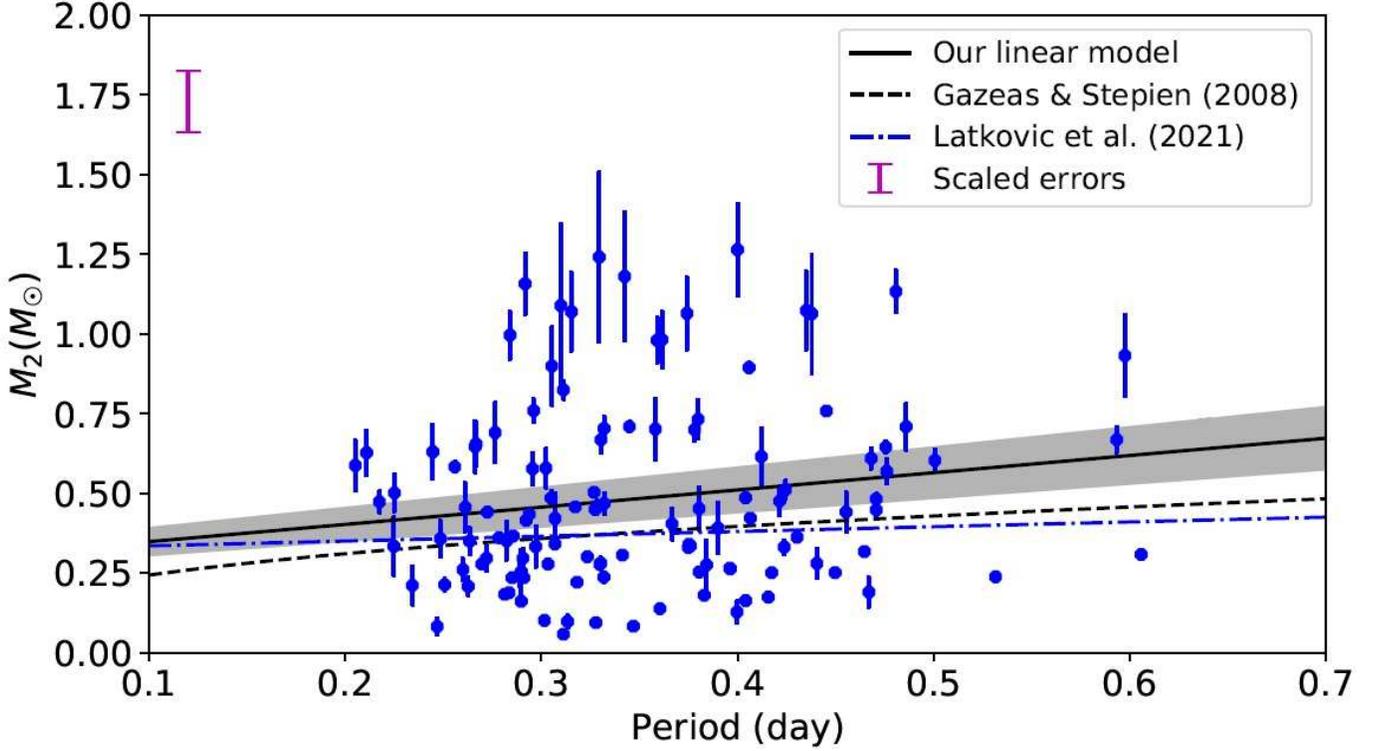}
\caption{Orbital period versus secondary star’s mass.}\label{Fig2}
\end{center}
\end{figure}

\begin{figure}
\begin{center}
\includegraphics[width=\columnwidth]{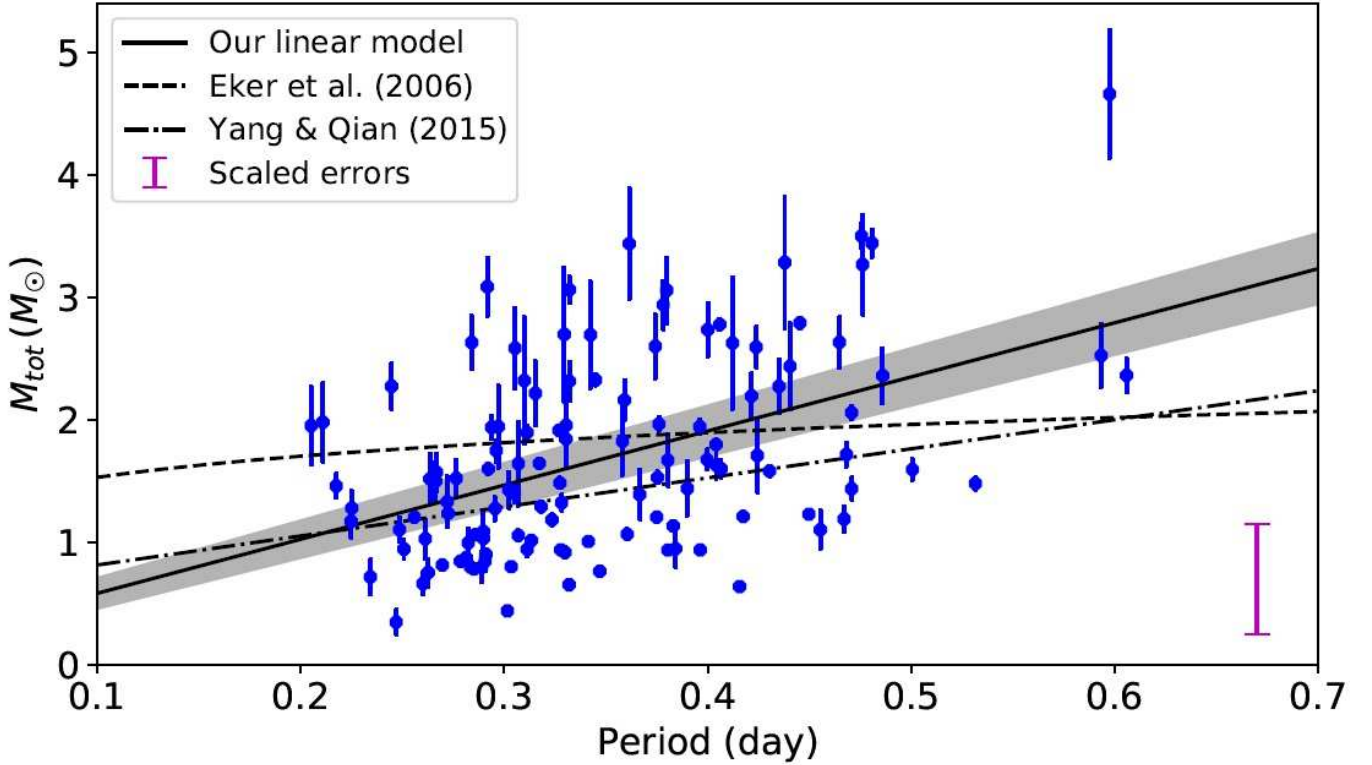}
\caption{Orbital period versus total mass of components.}\label{Fig3}
\end{center}
\end{figure}

We plotted the orbital period versus the mass of the components and the total mass based on Table \ref{tab} in Figures \ref{Fig1}, \ref{Fig2}, \ref{Fig3}. A linear model was fitted to the data points of these figures using the MCMC approach. This method was applied by sampling from the posterior probability distributions of the coefficients of the linear model (parameters of MCMC: 20 walkers, 5000 iterations and burn-in = 100) using the Pymc3 package in Python (\citealt{2015arXiv150708050S}).
The presented models of previous studies are shown in Figures \ref{Fig1}, \ref{Fig2}, \ref{Fig3} for comparison; in these figures, a linear (black line) model was applied to the data. The linear models show the median values, and shaded areas represent $1\sigma$ values (the intervals between the 16th and the 84th percentile values) for the model parameters. The blue circle marker shows the points derived from this study and the red circle marker shows the 86 points from the \cite{2018PASJ...70...90K} study.
The used error bar for all our data points during the linear model is represented by the magenta error bar; this is equal to the average uncertainty of our data points.

Since the error bars were underestimated for some of the data points, we used the average uncertainty of our data points $(\sigma)$ or 2$\sigma$ as a scaled error during the fits.

Based on the linear fit parameters, we updated the new relationships between orbital period and masses as:

\begin{equation}\label{eq9}
M_1=(2.924\pm0.075)P+(0.147\pm0.029)
\end{equation}

\begin{equation}\label{eq10}
M_2=(0.541\pm0.092)P+(0.294\pm0.034)
\end{equation}

\begin{equation}\label{eq11}
M_{tot}=(4.421\pm0.273)P+(0.138\pm0.099)
\end{equation}

\subsection{$P-T_1$ relationship}
Studies of the relationship between $P$ and $T_{\rm1}$ have been performed so far. In this regard, \cite{2011AJ....142..117S} investigated this relationship with a sample of 25 contact systems. $P-T_1$ relationship was also studied by \cite{2017RAA....17...87Q} for systems with an orbital period of less than 0.6 days. 
The relationship showed a large scatter caused by the effect of a third star within each system and the inaccuracy of some orbital periods. In another study, \cite{2020MNRAS.493.4045J} 
derived two linear fits from a large sample for this relationship (Table \ref{tab2}). Their diagram showed a break and divided samples into two classes. 
Early-type systems were analyzed to be hotter and had shorter orbital periods while late-type systems became hotter at longer orbital periods. The study 
by \cite{2020PASJ...72..103L} showed similar results to that of \cite{2017RAA....17...87Q}. \cite{2021ApJS..254...10L} investigated and derived a linear fit for the relationship between $P$ and $T_{\rm1}$. They reported a break in the $P-T_{\rm1}$ diagram similar to that of \cite{2020MNRAS.493.4045J}.

Table \ref{tab2} shows the results of the previous studies comprising measured orbital period-temperature relationships.

\begin{figure}
\begin{center}
\includegraphics[width=\columnwidth]{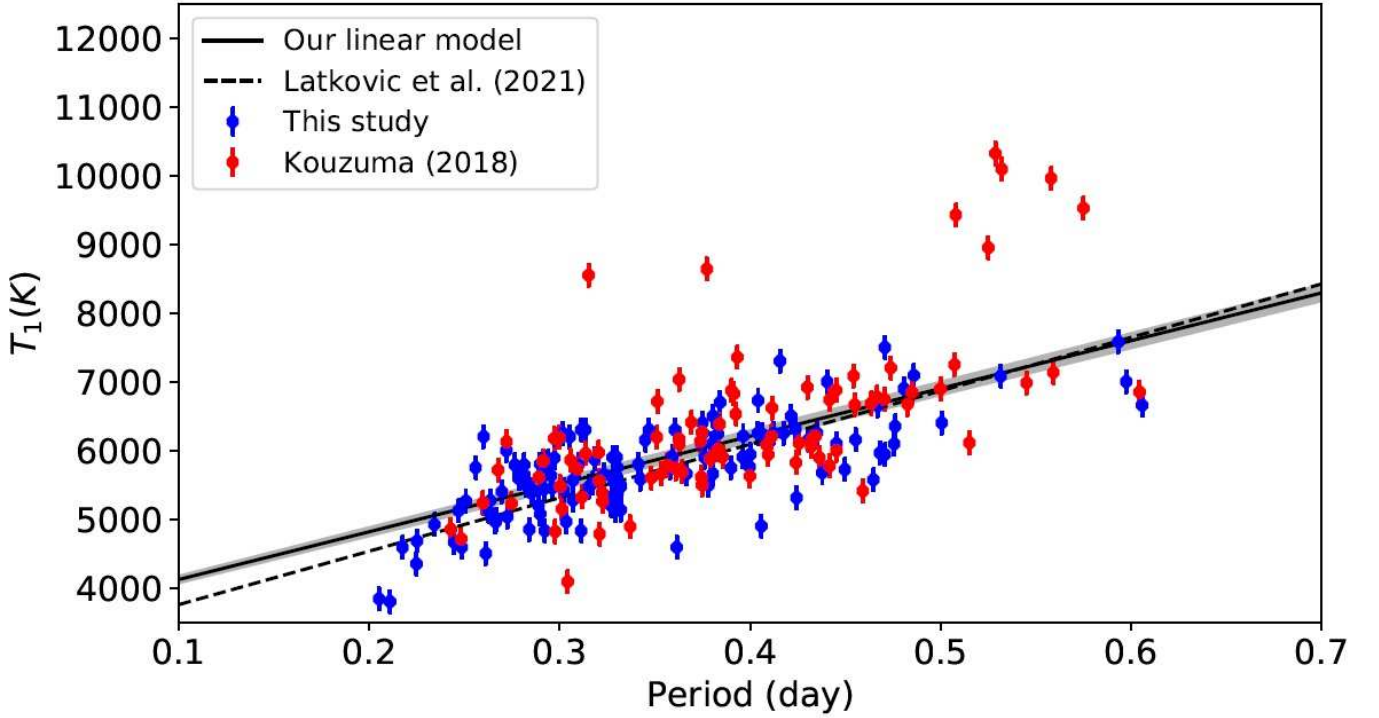}
\caption{Orbital period versus primary star’s temperature. A linear (black line) model was applied to the data. Shaded areas show the intervals 
between the 16th and the 84th percentile values for parameters of the linear model. The blue circle marker shows the points we derived and the red circle 
marker shows 86 points from the \protect\cite{2018PASJ...70...90K} sample.}
\label{Fig4}
\end{center}
\end{figure}

We drew the orbital period versus primary component’s temperature based on Table \ref{tab} in Figure \ref{Fig4}; the fit method is the same for 
all the figures in Section 4 as described in subsection 4.2. We derived the new relationship between them based on parameters of linear fit as:
\begin{equation}\label{eq12}
T_1=(6951.42 _{\rm-112.68}^{+112.16})P+(3426.01 _{\rm-43.90}^{+44.12})
\end{equation}

Figure \ref{Fig4} shows an obvious break after the 0.5 orbital period and the data does not follow the trend before it.

\begin{table*}
\caption{Relations between orbital period and temperature for EW-type systems based on previous studies.} 
\centering
\begin{center}
\footnotesize
\begin{tabular}{c c c }
 \hline
 \hline
        Parameters & Relation & Reference \\
	   \hline
       $P-T_{\rm1}$ & $T_{1}=4400(300)P+4025(640)$ & \cite{2011AJ....142..117S} \\ 
       $P-T_{\rm(late-type)}$ & $T=6598(\pm23)K+5260(\pm116)Klog(P/0.5 d)$ & \cite{2020MNRAS.493.4045J} \\ 
       $P-T_{\rm(early-type)}$ & $T=7041(\pm28)K-843(\pm164)Klog(P/0.5 d)$ & \cite{2020MNRAS.493.4045J} \\
       $P-T_{\rm1}$ & $T_1=(7780\pm307)P+(2977\pm105)$ & \cite{2021ApJS..254...10L}\\
       \hline
       \hline
\end{tabular}
\end{center}
\label{tab2}
\end{table*} 

\subsection{$P-log(g)$ relationships}
\cite{2017RAA....17...87Q} found that $log(g)$ is weakly correlated with orbital period in EW-type systems, and the systems with short-periods ($P<0.29$ d; $T<5700$ K) have a higher $log(g)$.

\begin{figure}
\begin{center}
\includegraphics[width=\columnwidth]{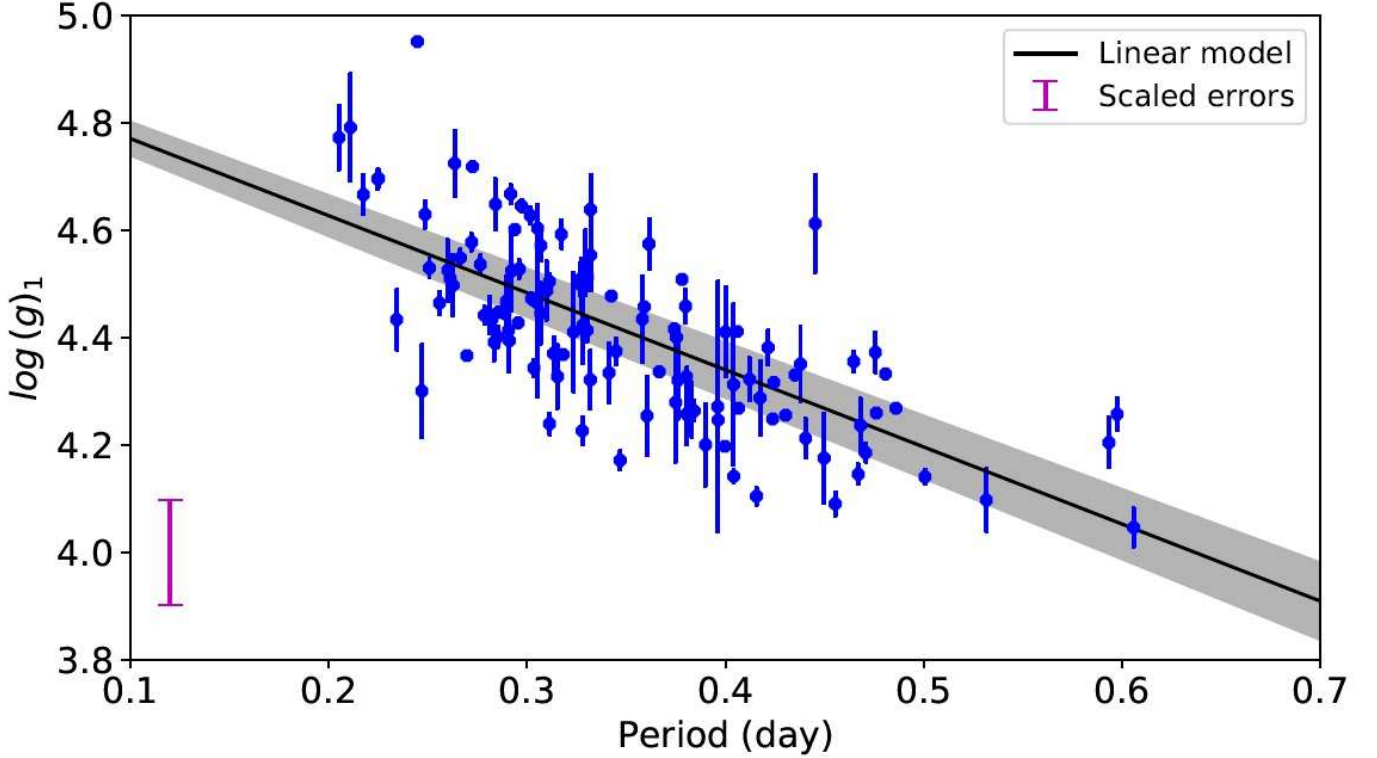}
\caption{Orbital period versus $log(g)$ of primary stars. The used error bar for all our data points in this model is represented by the magenta error bar in the lower left corner of the graph; this is equal to twice the average of the uncertainties of our data points.}\label{Fig5}
\end{center}
\end{figure}

\begin{figure}
\begin{center}
\includegraphics[width=\columnwidth]{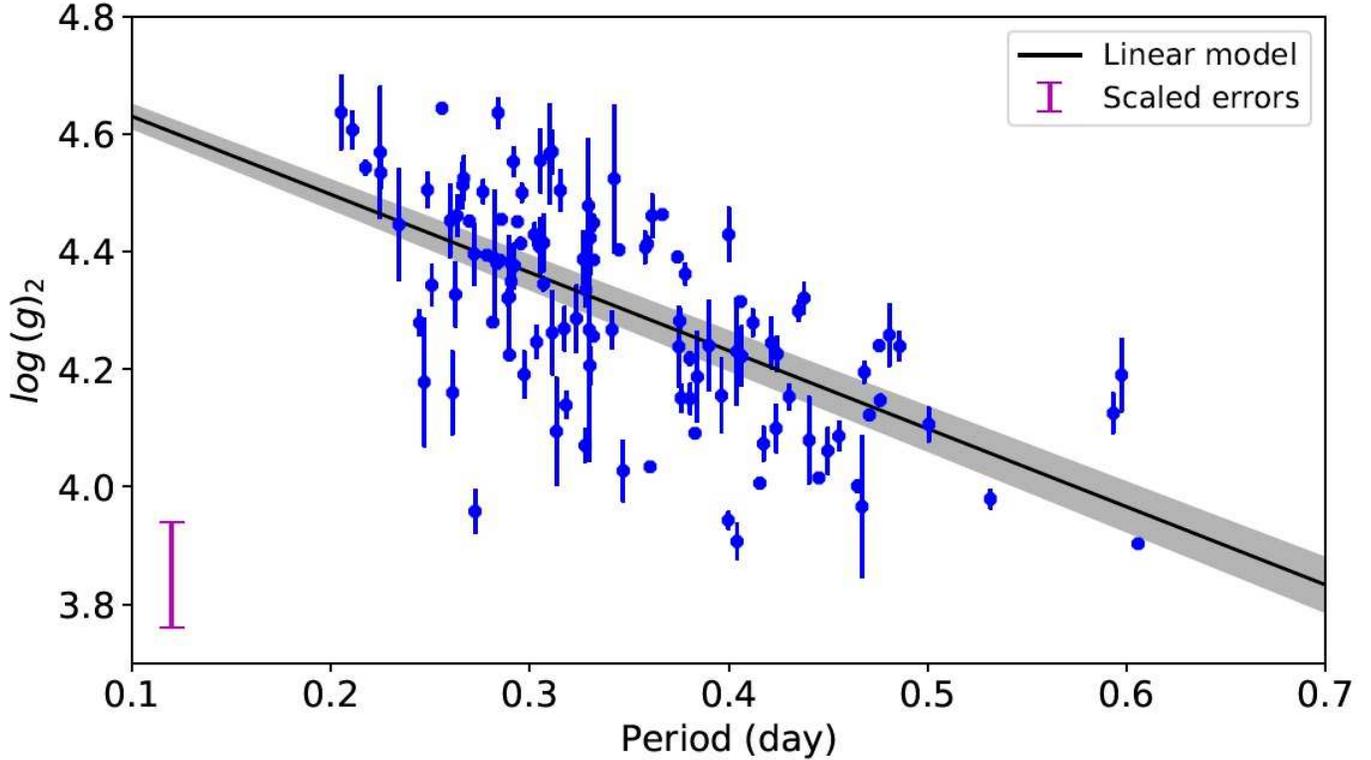}
\caption{Orbital period versus $log(g)$ of secondary stars. The used error bar for all our data points in this model is represented by the magenta error bar in the lower left corner of the graph; this is equal to the average of the uncertainties of our data points.}\label{Fig6}
\end{center}
\end{figure}

We drew the orbital period versus $log(g)$ based on Table \ref{tab} in Figures \ref{Fig5} and \ref{Fig6}. 
A linear (black line) model in each figure was applied to the data with $1\sigma$ values for the model parameters which were shown in the shaded areas. We derived the relationship between them based on parameters of linear fit as:
\begin{equation}\label{eq13}
log(g)_1=(-1.436\pm0.068)P+(4.914\pm0.025)
\end{equation}

\begin{equation}\label{eq14}
log(g)_2=(-1.329\pm0.044)P+(4.763\pm0.016)
\end{equation}

\subsection{$P-M_{\rm1}-T_{\rm1}$ relationship}
As discussed in this study, there is a relationship between $P-M_{\rm1}$ and $P-T_{\rm1}$. The relationship 
between $M_{\rm1}-T_{\rm1}$ in a sample of low-temperature contact binaries has been plotted by \citet{Yakut_2005}. However, most of the papers concentrating on the relationship between $M-T$ have studied samples of detached/semi-detached eclipsing binaries (e.g., \citealt{1967AcA....17....1P}; \citealt{1988BAICz..39..329H}; \citealt{2002ARep...46..233K}; \citealt{2007MNRAS.382.1073M};
\citealt{2013ApJ...776...87S}; \citealt{2018MNRAS.479.5491E}). 
Given the dual relationships that exist between the parameters ($P-M_{\rm1}$, $P-T_{\rm1}$, and $M_{\rm1}-T_{\rm1}$), we considered the relationship between these three parameters $P-M_{\rm1}-T_{\rm1}$ for contact systems.

\begin{figure*}[ht]
\begin{center}
\includegraphics[width=\columnwidth]{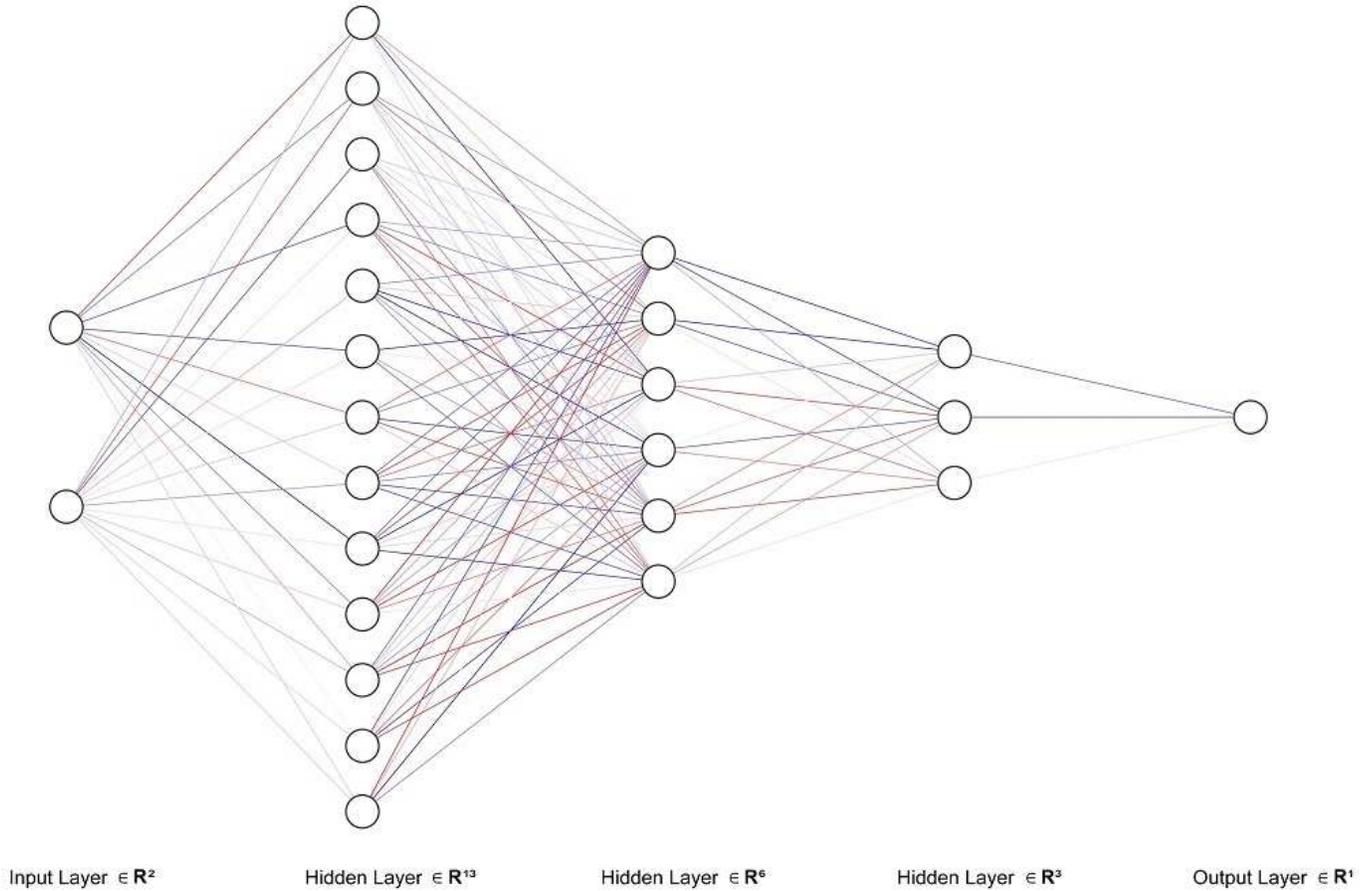}
\caption{The model architecture in the form of $N_{in}: N_{1}: N_{2}: N_{3}: N_{out}$ is represented as 2 : 13 : 6 : 3 : 1, 
where  $N_{in}$ is the number of input nodes, $N_{out}$ is the number of output nodes and  $N_{i}$ is the number of nodes 
in the $i^{th}$ hidden layer.}\label{Fig7}
\end{center}
\end{figure*}

In order to get the new orbital period-mass relationships for each component of the studied star systems, we implemented a machine learning model by a three-layer deep ANN using the Keras library on TensorFlow as the backend (\citealt{2016arXiv160304467A}). 
This ANN takes two inputs, which are the orbital period and the temperature of the primary component of the sample system, and yields the primary mass as its numerical output. The architecture of this fully connected sequential MLP is shown in Figure \ref{Fig7}. The number of hidden layers and their nodes were chosen based on previous experiments in order to prevent
over-fitting according to the limited size of our dataset. The training process was done by applying our dataset of 118 contact systems to the model; however, the validation set consisted of the dataset in \cite{2018PASJ...70...90K} including 86 systems. As we have designed a numerical computation model and required a quantitative output, the weights were initialized by the default normal kernel initializer (kernel\_initializer=’normal’). The activation function for all nodes in all layers was selected as the Rectified Linear Unit (ReLU) due to its good performance in numerical calculations (\citealt{Nair2010RectifiedLU}). 
Having the model created now, the next step was to define the model compilation parameters.

For the loss function, we preferred the well-known Mean Squared Error (MSE), because it is the most familiar method for error detection in statistics (\citealt{4775883}). This network will be evaluated by calculating the MSE of the training and validation datasets to measure its precision. Our optimizer was set to be the Adaptive Moment Estimation Algorithm (Adam) because it did well in a close competition among the other optimization algorithms and also because Adam was suggested for models with numerical output 
(\citealt{2014arXiv1412.6980K}). After compilation, the model needs to be evaluated. To do so, by trial and error the epoch was set to be 500 and the batch size to 5. The low number of samples in our dataset caused over-fitting after approximately 250 epochs. As a result, the loss function did not decrease considerably, and the model showed no significant difference in the cycles of optimization.

\begin{figure}
\begin{center}
\includegraphics[width=\columnwidth]{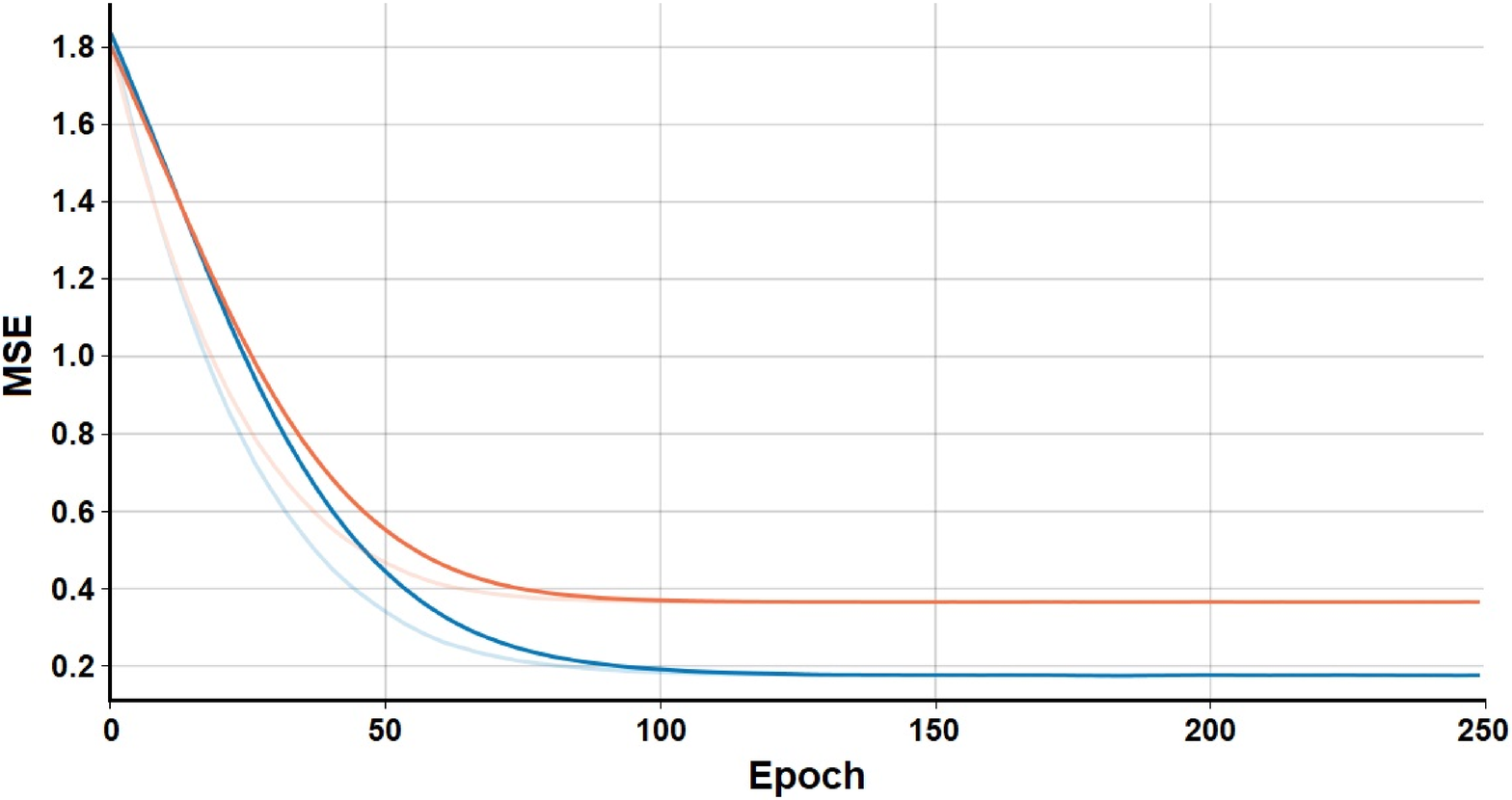}
\caption{The value of MSE in terms of the number of epochs for the training set (orange line) and the validation set (blue line). The curves with the brighter colors show the smoothness with a different slope.}\label{Fig8}
\end{center}
\end{figure}

The relation of MSE in terms of the number of epochs (MSE-Epoch) is represented in Figure \ref{Fig8}. For the training set, the diagram shows a sharp drop in the error value from 1.8 to almost 0.48 until 50 epochs; then after 100 times of training, the value of MSE decreases smoothly to nearly 0.38 and gradually tends to zero. The validation set demonstrates an almost similar trend and even after approximately 20 epochs shows a higher error reduction to about 0.18, which is a sign that the training process 
has been done well. In order to be more precise, we took advantage of K-fold Cross Validation (CV), k=10, to create a training network. 
Training and testings were done 10 random times using cross\_val\_score each time in the iteration loop; then the mean of outputs was calculated to get the best prediction result.

The outputs of this method can help to estimate the primary companion mass. It should be noted that this estimate can be made even before performing light curve solutions. The orbital period of the system is obtained in a study of orbital period changes, and the temperature of the primary star is also generally fixed in studies, which is itself the result of conventional methods such as ($B-V$) or Gaia DR2 results.

\subsection{$P-q$ relationship}
Some investigations used different models in order to find out whether there is a relationship between the mass ratio and the orbital period. \cite{2015ApJ...810...61M} used the Optical Gravitational Lensing Experiment (OGLE-III) database to measure binary properties and mass ratio. In another paper published by \cite{2017ApJS..230...15M}, compiling observations of early-type binaries, they attempted more to investigate  the interrelation between the orbital period and the mass ratio distribution.

In order to study the characteristics of W UMa contact binaries, \cite{2020MNRAS.497.3493Z} collected data for 369 contact binaries and fitted polynomials to their diagrams. By combining their obtained best-fit equations with Kepler's third law, they presented a $P-a$, and $P-q$ relation. It should be noted that the $q$ defined in their paper is the ratio of the less massive star mass to the more massive star mass, and therefore $q$ is always less than unity. We investigated their collected data thoroughly and improved it by removing the repeated objects and objects with less certain values of $q$. In our investigation, we selected binary systems that have light curve analyses using both photometric and spectroscopic results. If the selected system has only photometric analysis, we have chosen the most recent study of that system. We eventually used the 251 systems from a study by \cite{2020MNRAS.497.3493Z} where we found referenced papers and added our sample in this study to the dataset.

\begin{figure}
\begin{center}
\includegraphics[width=\columnwidth]{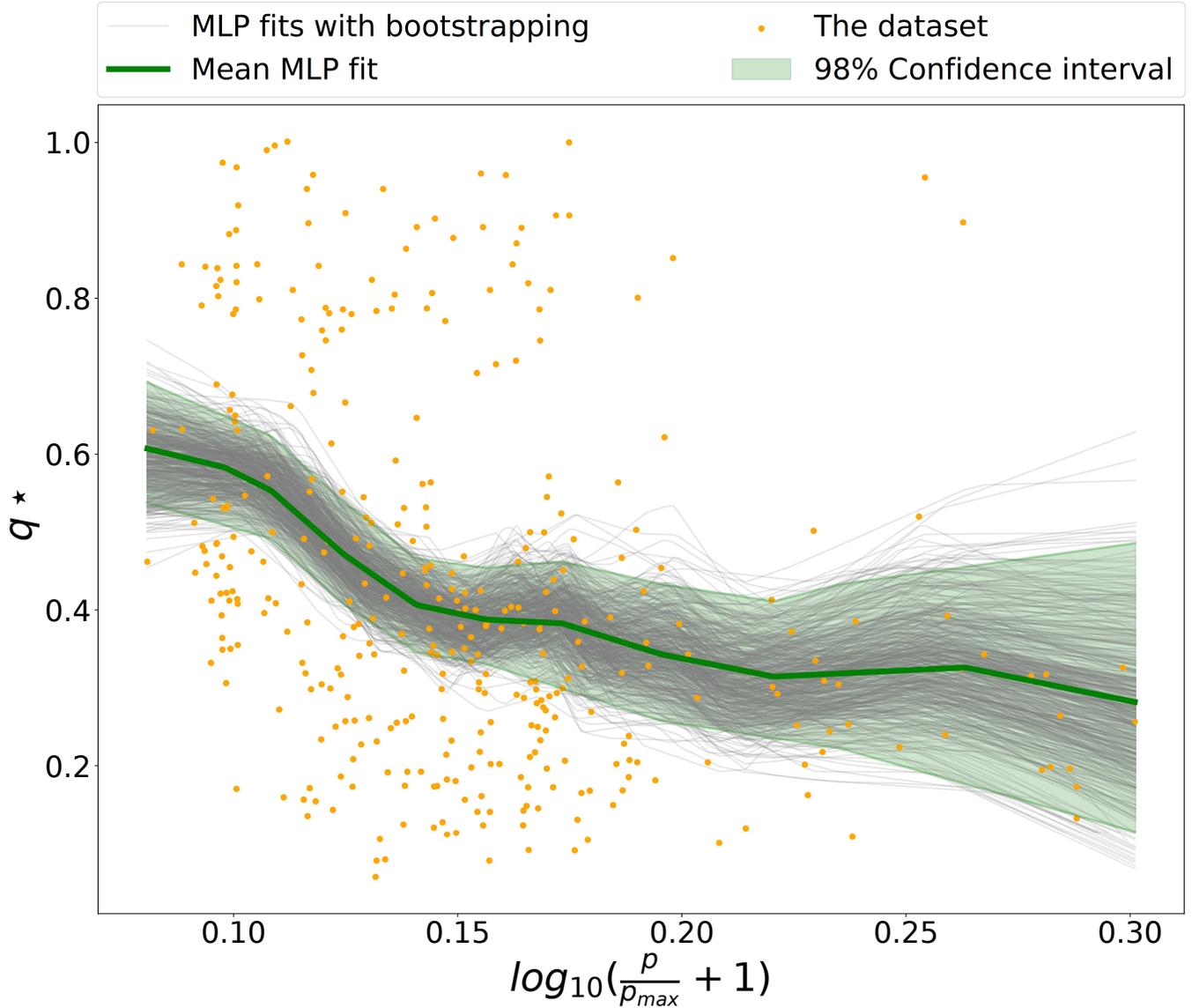}
\caption{This figure shows the plot of the ratio of the less massive star to the more massive star ($q^*$) 
versus orbital period in units of days. The gray curves represent 500 MLP regressors fitted to randomly selected subsets of the data independently. 
The green curve shows the mean MLP model and the green shaded area represents the 98\% confidence interval of all 500 fits.}\label{Fig9}
\end{center}
\end{figure}

This enhanced dataset was then used to obtain a new relation between orbital period and mass ratio. We define the mass ratio similar to 
\cite{2020MNRAS.497.3493Z}, and hereafter, we call it $q^*$ to avoid any confusion. In order to use the orbital periods as the input to our regressor, we first normalize them by dividing all values by the maximum value. Next, we calculate  $log_{\rm10} (p+1)$ and use it as our input. The new relation is found by fitting a MLP regression model\footnote{Parameters of the MLP regressor: Number of hidden layers=2, Size of the layer=(30,10), Batch size=64, Activation function= The rectified linear unit function, Tolerance for the optimization$=1\times10^{(-7)}$, Maximum number of iterations=2000.}, and the confidence interval of the fit is obtained by bootstrapping the fits in 500 steps. For each 500 steps, 90\% of the data is randomly chosen with replacement, and the model is applied to them. Figure \ref{Fig9} shows the fitted model along with the data.
The gray curves show the fitted model at each step. The green curve is the mean of the 500 fitted models, and the green shaded area represents the 98\% confidence interval of the fits. Note that the data here has been collected from many dependent investigations. Therefore, we have fitted the 500 models to the random subsets of the data to account for this wide diversity in the data and detect the most significant trend; The scatter in the models (grey curves) is due to this diversity. As expected, the final model has a higher uncertainty in the regions with a smaller number of objects. \cite{2020MNRAS.497.3493Z} estimates an extremely large range for the mass ratio of the systems, and our introduced $P-q^*$ relation significantly enhances these estimations. Such a method has a high degree of flexibility and allows investigating various possible models.

\section{Summary and conclusion}\label{sec5}
We computed the mass and other absolute parameters of 118 contact binaries ($P<0.6$ d) using the Gaia EDR3 parallaxes and parameters of light curve solutions from previous studies. Then, we analyzed and produced $P-M_{\rm(1,2)}$, $P-log(g)_{\rm(1,2)}$, and $P-T_{\rm1}$ 
relations. The MCMC method has been utilized for the relations for which we updated the formula. Also, the MLP and ANN methods were used to investigate $P-q$ and $P-M_{\rm1}-T_{\rm1}$ relationships, respectively. In all steps and relationships, we have considered the more massive component as the primary star. The previous investigations used photometric light curves and radial velocity measurements mostly derived from 
the literature to obtain absolute parameters (\citealt{2003MNRAS.342.1260Q}, \citealt{2006MNRAS.373.1483E},
\citealt{2008MNRAS.390.1577G}, \citealt{2015AJ....150...69Y},
\citealt{2021ApJS..254...10L}). 
However, this is the first study to apply the Gaia parallax method to determine the masses of stars to evaluate current relationships in these types of systems.

The new relationship between $P$ and $M_{\rm1}$ was presented in this study, and it shows that the mass reduces as the orbital period decreases. According to the \cite{2018PASJ...70...90K} study, the rate of mass transferring from the more-massive to the less-massive component of a contact system increases as the binary evolves. Furthermore, the relationship between $P$ and $M_{\rm2}$ has been of interest in previous studies, such as \cite{2003MNRAS.342.1260Q}, \cite{2006AcA....56..199S}, \cite{2006MNRAS.370L..29G}, \cite{2018PASJ...70...90K}. \cite{2021ApJS..254...10L} also reported a relationship between the 
dependency of $M_{\rm2}$ on $P$. As a result, despite the fact that $P-M_{\rm2}$ has a higher scattering rate than $P-M_{\rm1}$, 
we concluded that $P$ and $M_{\rm2}$ have a relationship.

The relationship between the orbital period and the temperature can also be significant according to studies by \cite{2017RAA....17...87Q}, 
\cite{2020MNRAS.493.4045J}, \cite{2020PASJ...72..103L}, and \cite{2021ApJS..254...10L}. Hence, we provided an updated relationship with the sample and derived a linear fit for the correlation between the orbital period and the primary temperature. We achieved the same result as 
\cite{2021ApJS..254...10L} and \cite{2020MNRAS.493.4045J} validating a clear break in the orbital period longer than 0.5 days.

We also investigated the relationship between the orbital period and $log(g)$ in our sample. Accordingly, as the orbital period increases, the value of the $log(g)$ decreases. Therefore, it can be concluded that there is a significant relationship between the orbital period and $log(g)$ parameters in contact systems. Due to the scattering of points, this significant relationship is especially strong in the case of the $P-log(g)_{\rm1}$ relationship.

According to the relationship between $P-M_{\rm1}$ and $P-T_{\rm1}$, we examined the interaction between the $P-M_{\rm1}-T_{\rm1}$ parameters. Based on this, we presented a model that is the result of data analysis through the ANN method. This model shows that there 
is a significant interaction between these three parameters simultaneously. Therefore, $M_{\rm1}$ can be estimated if one knows both the orbital period and the temperature of the primary component. Certainly, with a larger sample, this model will improve with the ANN method. We selected seven systems to compare the outputs of the ANN method and equation 9 with other studies' results. These systems were not present in the samples used in this study, and the primary component mass was determined through spectroscopy. According to Table \ref{tab3}, the ANN method's results appear to be capable of providing an acceptable estimate.

As it can be seen from Table \ref{tab3}, the value of $M_{1}$ using equation 9 also had acceptable results.
In addition, the total mass obtained in this study based on the Gaia EDR3 parallax, compared to the total mass from the literature that used the photometric method, shows about 18\% improvement over spectroscopic investigations.

\begin{table*}
\caption{A comparison between the ANN method and equation 9's outputs, with previous spectroscopy-based investigations for the primary component masses.} 
\centering
\begin{center}
\footnotesize
\begin{tabular}{c c c c c c c c }
 \hline
 \hline
         System & P(day)& $T_{1}(K)$ & $M_{1}(M_{\odot})$& Reference & $M_{1}(M_{\odot})$ by ANN & $M_{1}(M_{\odot})$ by Eq. 9\\
		\hline
        V546 And & 0.3830195(43)&6517 &1.083(30)&\cite{2015NewA...36..100G}& 1.107(34) & 1.267(58)\\
        GM CVn & 0.3669835(1)& 5920&1.23(6) & \cite{2018JAVSO..46...43A} & 1.216(38) & 1.220(57)\\
        DX Tuc&0.37711010(2)&6250& 1.00(3)&\cite{2007AandA...465..943S} & 1.063(36) & 1.250(57)\\
        EH Cnc& 0.4180358(2)& 6390& 1.28(2)& \cite{2020JApA...41...26A} & 1.349(49) & 1.369(54)\\
        V1918 Cyg& 0.4131769(2)& 7300& 1.302(69)&\cite{2016NewA...47...57G} & 1.222(35) & 1.355(60)\\
        TYC 1174-344-1& 0.3887124(1)& 6500&1.381(14)&\cite{2011NewA...16..242G} & 1.145(35) & 1.284(58)\\
        \hline
        \hline
\end{tabular}
\end{center}
\label{tab3}
\end{table*}

Finding a relation between the mass ratio and the orbital period of W UMa binaries has been an interesting topic for astronomers in this field for a long time. We studied the interactions found in the literature and aimed at enhancing them with new techniques. \cite{2020MNRAS.497.3493Z} presented one of the most recent models, and we investigated the data from their analysis and enhanced it by removing some of the objects and adding our sample. We then presented a new model between the mass ratio and the orbital period by bootstrapping an MLP regression model in 500 steps. It is important to note that in the case of mass ratio and orbital period, there is no true model to look for, and our goal here is to test innovative regression models to probe different aspects of the data. This is an initial step to delve more deeply into the details of this problem. In the next steps, we will test this model on a larger number of events and will try to include more parameters or augment the parameters that we already have by including various combinations of them. This would allow the model to detect patterns that exist in different parameter spaces. The model presented here can make decent predictions for the mass ratio of the W UMa systems based on their orbital period with lower errors and a mean squared error of 0.046.

The relationships between the parameters can reveal information about the systems’ evolution. They could be useful for determining the contact binary parameters based on the orbital period.

\begin{figure}
\begin{center}
\includegraphics[width=\columnwidth]{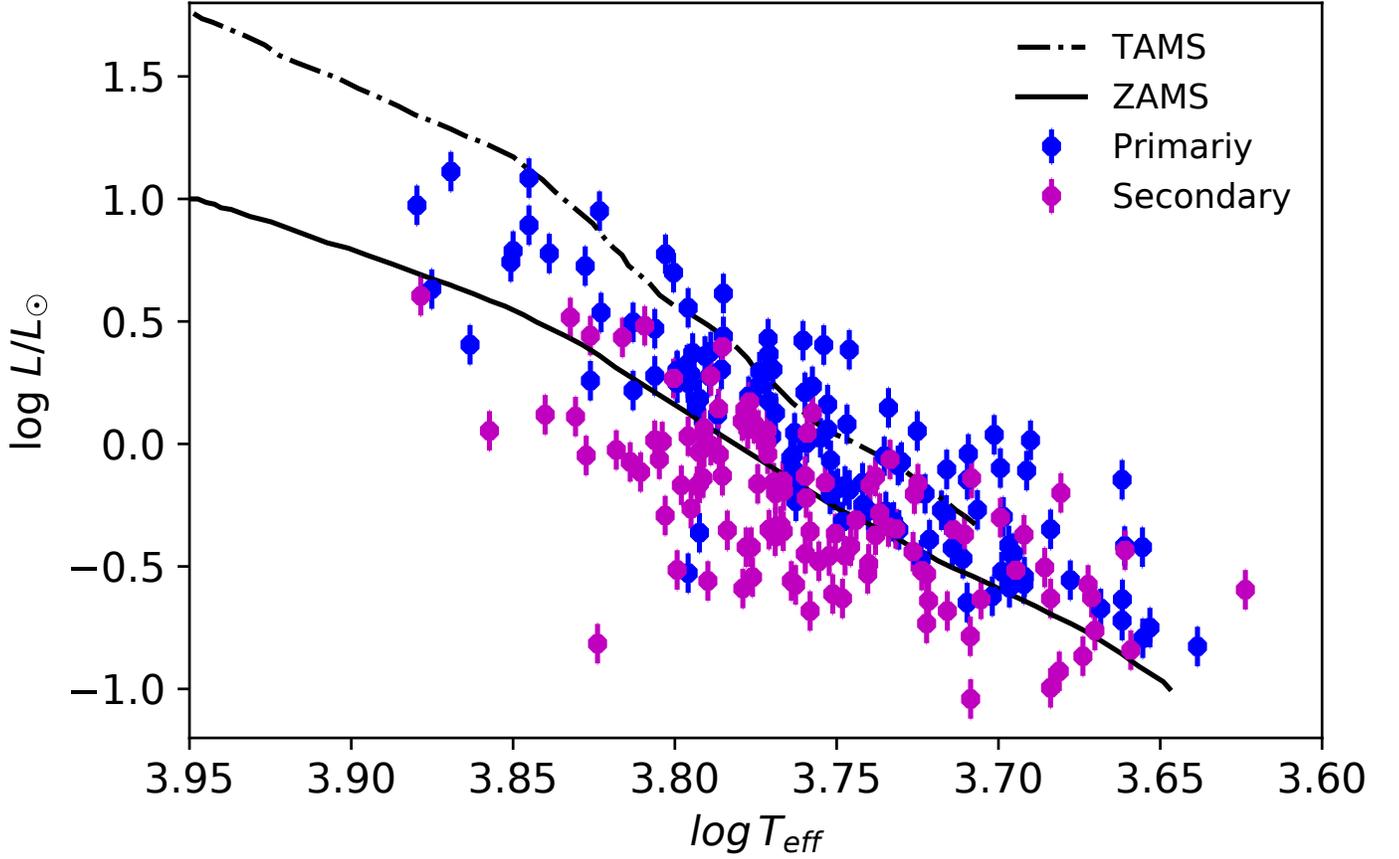}
\caption{The positions of the primary and secondary stars are shown in the H-R diagram. Values of 
luminosity were obtained by calculating the absolute parameters in this study. The average uncertainty 
of the luminosity was estimated by a Monte Carlo process.}\label{Fig10}
\end{center}
\end{figure}

The components of the contact systems are plotted in the Hertzsprung-Russell (H-R) diagram (Figure \ref{Fig10}).
Most of the primary stars are found between Zero Age Main Sequence (ZAMS) and Terminal Age Main Sequence (TAMS). In addition, the secondary stars are closest to and below ZAMS. The location of both components of each system appears to be in good agreement with the distribution of primary and secondary stars of the W UMa systems.\\

\begin{figure}
\begin{center}
\includegraphics[width=\columnwidth]{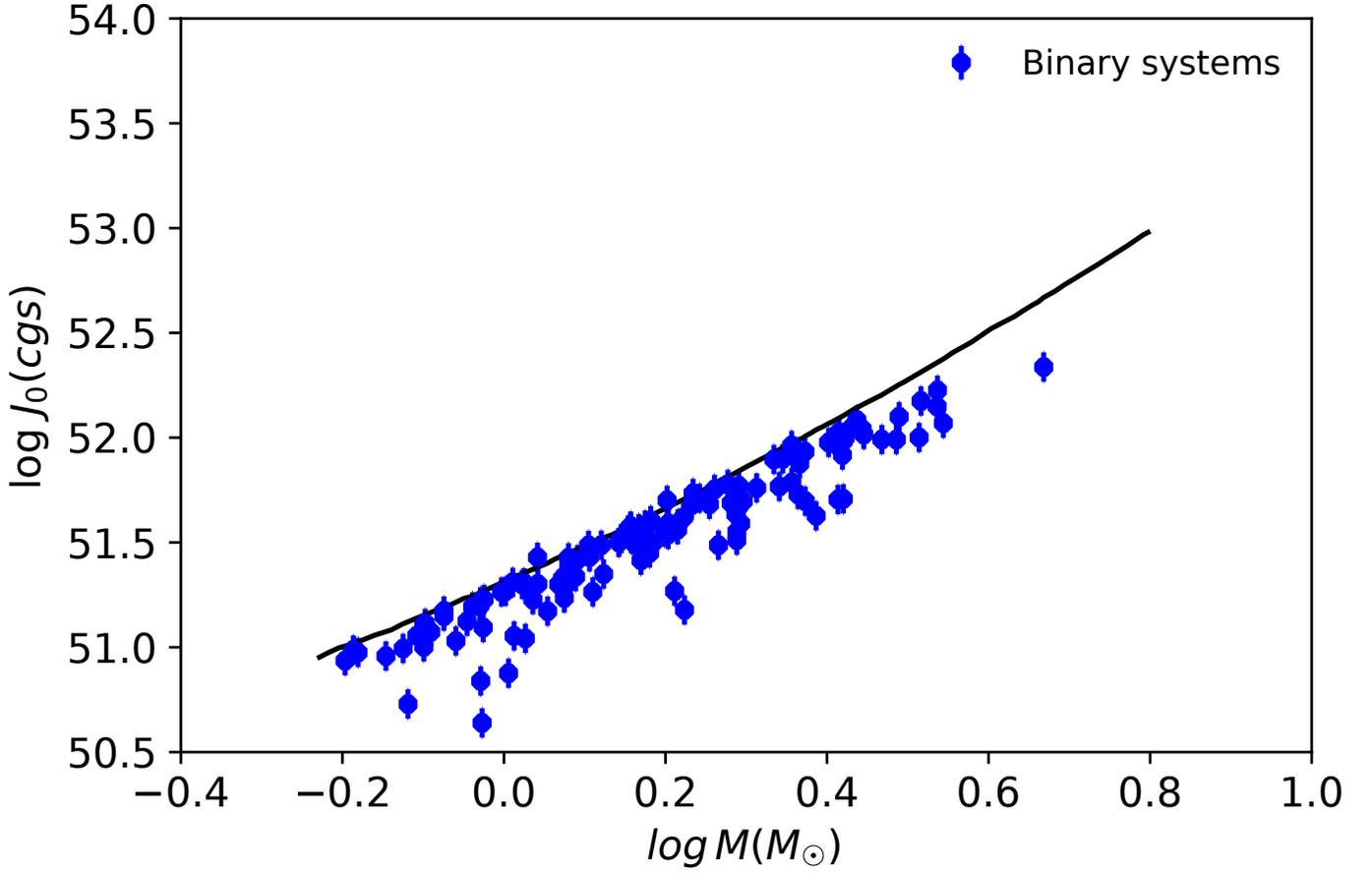}
\caption{Location of the 118 systems on the $logJ_0-logM$ diagram. Detached binaries are above the quadratic border line, and contact systems are below it (\citealt{2006MNRAS.373.1483E}).}\label{Fig11}
\end{center}
\end{figure}

We estimated our 118 systems' Orbital Angular Momentum ($J_0$), and their locations are shown on the $logJ_0-logM$ diagram (Figure \ref{Fig11}). For this computation, we utilized the following equation from \cite{2006MNRAS.373.1483E}, which is the result of a quadratic line that separates detached and contact systems.

\begin{equation}\label{eq15}
J_0=\frac{q}{(1+q)^2}\sqrt[3]{\frac{G^2}{2\pi}{M^5}{P}}
\end{equation}

where $q$ is the mass ratio, $M$ is the total mass of the system, $P$ is the orbital period and $G$ is the gravitational constant. As a consequence, $J_0$ was found to be below the quadratic curve and in the contact binary region in all of our systems.

It is critical to select reliable samples for investigating relationships. We tried to achieve the required uniformity in sample selection. Although the observation tools, data reduction techniques, and analysis methods may differ, we tried to reduce the effect of these variations by choosing the correct quantity of data and flexible methods.

\section{Data Availability}\label{sec6}
The model used about the $P-q^*$ relationship is available at:
\url{https://github.com/Somayeh91/p\_q\_relations\_contact\_binaries}

In addition, for convenience of usage, the model used for the $P-M_{\rm1}-T_{\rm1}$ relationship 
has been provided at this address:
\url{https://astronomy.raderonlab.ca/AIModels}

It should be noted that these models will be updated based on the new results.

\section{Acknowledgments}\label{sec7}
This work has made use of data from the European Space Agency (ESA) mission Gaia from which we used the latest version Gaia EDR3 (\url{http://www.cosmos.esa.int/gaia}). We also utilized the SIMBAD database and VizieR by the APASS9 and ASAS catalogs, operated at the CDS center, Strasbourg, France (\url{https://vizier.u-strasbg.fr/viz-bin/VizieR-4)}.

Furthermore, we would like to thank Afshin Oroojlooyjadid for his instructive suggestions and contributions regarding the machine learning approaches implemented in this work. We express our gratitude to Shinjirou Kouzuma and Edwin Budding for their scientific advice. Also, our sincere thanks to Paul D. Maley for editing the text.

\clearpage
\appendix
\label{sec:appendix}

Parameters of previous studies:\\
Parameters obtained from the light curve solutions of the previous studies for the calculation of absolute parameters based on Gaia EDR3 are presented in this section.

\begin{table*}
\caption{Some previous studies' light curve analysis results.} 
\centering
\begin{center}
\footnotesize
\begin{tabular}{c c c c c c c c }
 \hline
 \hline
         System & $P$(day) & $l_{1}/l_{tot}(V)$ & $T_{1}(K)$ & $T_{2}(K)$ & $q$ & Reference\\
		\hline
1&
0.37602172(6)&
0.7962(8)&
5900&
5882(4)&
0.2055(8)&
\cite{2019JASS...36..265K}\\
2&
0.30231846(5)&
0.464(4)&
5420(6)&
5201&
1.460(14)&
\cite{2021RAA....21..203P}\\
3&
0.404160442(752)&
0.8978&
6725&
6185(29)&
0.111(2)&
\cite{2020NewA...7901391O}\\
4&
0.32690440000(2)&
0.7714(128)&
5200(17)&
4950(16)&
0.3572(9)&
\cite{2020RoAJ...30..189A}\\
5&
0.4238006(5)&
0.863(16)&
6315&
6114(15)&
0.1464(17)&
\cite{2016NewA...44...35B}\\
6&
0.468126612(66)&
0.3767(13)&
5964(5)&
5920&
1.8157(88)&
\cite{2011PASP..123...34L}\\
7&
0.33200448(3)&
0.3116(16)&
5470&
5005(11)&
4.244(42)&
\cite{2018PASP..130g4201L}\\
8&
0.26125919(3)&
0.4601(260)&
4500&
3949(28)&
0.800&
\cite{2018NewA...59....1M}\\
9&
0.291740(6)&
0.67109(1)&
5436&
5005(1)&
0.600(4)&
\cite{2018NewA...64....1L}\\
10&
0.296096(9)&
0.49822(4)&
5454&
5219(6)&
1.300(5)&
\cite{2018NewA...64....1L}\\
11&
0.38040619(2)&
0.7076(32)&
5660&
5669(13)&
0.371(5)&
\cite{2016NewA...43....1L}\\
12&
0.41216837(9)&
0.770(10)&
6250&
6180(5)&
0.306(9)&
\cite{2011NewA...16..194R}\\
13&
0.378141051(88)&
0.3616(17)&
5492&
5027&
3.200&
\cite{2020RAA....20...96S}\\
14&
0.266350(9)&
0.641(5)&
4959&
4690(30)&
0.76(2)&
\cite{2020OAst...29...72D}\\
15&
0.276476955(57)&
0.621(8)&
5789&
5392(35)&
0.831(17)&
\cite{2021NewA...8401499R}\\
16&
0.2486159&
0.619&
4590&
4720&
0.48(5)&
\cite{2016AJ....152...57D}\\
17&
0.2447151(8)&
0.397(27)&
4660&
4204(6)&
0.38396(480)&
\cite{2016NewA...46...25M}\\
18&
0.295516571(23)&
0.481(6)&
5634&
5949(39)&
0.829(14)&
\cite{2021NewA...8401499R}\\
19&
0.33200799(3)&
0.3169(11)&
5135&
4913(5)&
3.353(13)&
\cite{2018PASP..130g4201L}\\
20&
0.2814114(1)&
0.24505(93)&
5790&
5649(6)&
3.7729(70)&
\cite{2016NewA...46...31G}\\
21&
0.21095&
0.3517&
3800&
3759&
2.154(74)&
\cite{2015AJ....149..111L}\\
22&
0.6057708(10)&
0.8736(7)&
6657&
6015(9)&
0.1499(3)&
\cite{2021NewA...8401512M}\\
23&
0.3960680(2)&
0.8164(21)&
5905&
5840(11)&
0.1579(8)&
\cite{2021NewA...8401512M}\\
24&
0.4646987(12)&
0.8007(17)&
5573&
5321(4)&
0.1371(11)&
\cite{2021NewA...8401512M}\\
25&
0.3469788&
0.8655(4)&
6300&
6163(9)&
0.122&
\cite{2004AJ....128.2430Q}\\
        \hline
\end{tabular}
\end{center}
\label{tab9}
\end{table*}

\begin{table*}
\renewcommand\thetable{tab9}
\caption{Continued} 
\centering
\begin{center}
\footnotesize
\begin{tabular}{c c c c c c c c }
 \hline
 \hline
         System & $P$(day) & $l_{1}/l_{tot}(V)$ & $T_{1}(K)$ & $T_{2}(K)$ & $q$ & Reference\\
		\hline
26&
0.32783433&
0.8514(2)&
5900&
6012(13)&
0.1115(3)&
\cite{2005MNRAS.356..765Q}\\
27&
0.44528226&
0.3108(49)&
6100&
5987 (26)&
2.68097&
\cite{2007AJ....134.1769Q}\\
28&
0.3749180&
0.6869(6)&
6400&
6513(10)&
0.3782(11)&
\cite{2008PASJ...60...77Q}\\
29&
0.406543(4)&
0.6923(1)&
6230(267)&
6366(3)&
0.359(1)&
\cite{2020JAVSO..48...40A}\\
30&
0.255886072(69)&
0.514&
5750(90)&
5180&
1.062(1)&
\cite{2015NewA...37...64T}\\
31&
0.440416842(9)&
0.8560(18)&
7000&
6920(2)&
0.130&
\cite{2016ApJ...817..133Z}\\
32&
0.31124849(8)&
0.6023(9)&
4830&
4702(2)&
0.773&
\cite{2013AJ....145...39Z}\\
33&
0.41567758(8)&
0.7066(8)&
7300&
7200(7)&
0.376(1)&
\cite{2003AJ....126.1960Y}\\
34&
0.26350684(15)&
0.3864(27)&
5268(19)&
4762&
3.331(52)&
\cite{2018PASJ...70..104X}\\
35&
0.27205&
0.3183(26)&
5998(17)&
5573&
3.493(63)&
\cite{2018PASJ...70..104X}\\
36&
0.311329(8)&
0.9277(28)&
6300&
6667(38)&
0.0652(13)&
\cite{2016AJ....151...69S}\\
37&
0.2600776&
0.637&
6200&
5970(20)&
0.65(8)&
\cite{2016AJ....152...57D}\\
38&
0.2248416&
0.584&
4350&
4800(20)&
0.40(4)&
\cite{2016AJ....152...57D}\\
39&
0.361456&
0.699&
4590&
4580(20)&
0.40(5)&
\cite{2016AJ....152...57D}\\
40&
0.27251563(4)&
0.186(3)&
5034&
4794(8)&
0.5555(2)&
\cite{2017NewA...57...37B}\\
41&
0.285702900(2)&
0.5880(2)&
5430&
5695(1)&
0.528(1)&
\cite{2018AcA....68..449A}\\
42&
0.2910143(2)&
0.6875(1)&
5620&
5807(2)&
0.354(1)&
\cite{2018AcA....68..449A}\\
43&
0.2851393(3)&
0.6388(1)&
5380&
5600(1)&
0.431(1)&
\cite{2018AcA....68..449A}\\
44&
0.26966078(2)&
0.6389(3)&
5400&
5639(3)&
0.425(1)&
\cite{2018AcA....68..449A}\\
45&
0.2834962(1)&
0.7092(2)&
5520&
5730(2)&
0.309(1)&
\cite{2018AcA....68..449A}\\
46&
0.3302577(2)&
0.786(2)&
5752&
5628(10)&
0.1799(16)&
\cite{2018NewA...62..121M}\\
47&
0.290627831(235)&
0.5860&
5789(9)&
6078&
0.5404(51)&
\cite{2019NewA...71...25O}\\
48&
0.399986(2)&
0.532&
5940&
5983(16)&
0.86(3)&
\cite{2014ApSS.353..575P}\\
49&
0.309898(2)&
0.717&
5794&
4921(41)&
0.886(15)&
\cite{2014ApSS.353..575P}\\
50&
0.359082527(18)&
0.502(2)&
5909&
5688(15)&
1.203(7)&
\cite{2019RAA....19...99M}\\
        \hline
\end{tabular}
\end{center}
\label{tab9}
\end{table*}

\begin{table*}
\renewcommand\thetable{tab9}
\caption{Continued} 
\centering
\begin{center}
\footnotesize
\begin{tabular}{c c c c c c c c }
 \hline
 \hline
         System & $P$(day) & $l_{1}/l_{tot}(V)$ & $T_{1}(K)$ & $T_{2}(K)$ & $q$ & Reference\\
		\hline
51&
0.31537&
0.572(4)&
5470&
5120&
1.071(10)&
\cite{2019RAA....19...99M}\\
52&
0.2938251(48)&
0.258(5)&
5841&
5745(39)&
3.461(22)&
\cite{2019RAA....19...99M}\\
53&
0.4378435(2)&
0.650(1)&
5676&
5721(6)&
0.479(7)&
\cite{2020RAA....20..179G}\\
54&
0.34134713(36)&
0.690&
5800(4)&
5730(4)&
0.439(1)&
\cite{2015NewA...34..262H}\\
55&
0.366599&
0.6960(29)&
5660&
5653(11)&
0.41(1)&
\cite{2018NewA...65...52H}\\
56&
0.2667676(1)&
0.4727(44)&
5000&
4840(13)&
1.396(28)&
\cite{2018NewA...61....1K}\\
57&
0.329520(6)&
0.372(15)&
5124(9)&
5743(177)&
0.853(13)&
\cite{2020NewA...8101439D}\\
58&
0.317260(4)&
0.6468(3)&
5521(168)&
5842(5)&
0.386(1)&
\cite{2020JAVSO..48...40A}\\
59&
0.217513&
0.6637&
4589&
4561(4)&
0.4800(27)&
\cite{2017NewA...50...37D}\\
60&
0.33189213&
0.453&
5500&
5140(10)&
1.751&
\cite{2004AcA....54..195B}\\
61&
0.30500298&
0.428&
6200&
5810(25)&
1.880&
\cite{2004AcA....54..195B}\\
62&
0.32990441&
0.373&
5900&
5590(33)&
2.300&
\cite{2004AcA....54..195B}\\
63&
0.47069442&
0.808&
7500&
6700(64)&
0.306&
\cite{2004AcA....54..195B}\\
64&
0.2785137(5)&
0.512(22)&
5590&
5295(8)&
1.3362(42)&
\cite{2018NewA...58...90B}\\
65&
0.4553331(2)&
0.616&
6156(35)&
5991(25)&
0.67(12)&
\cite{2021AJ....161..221P}\\
66&
0.246983(2)&
0.746&
5125(41)&
5112(3)&
0.31(1)&
\cite{2021AJ....161..221P}\\
67&
0.234278(1)&
0.706&
4921(51)&
4830(3)&
0.42(1)&
\cite{2021AJ....161..221P}\\
68&
0.38043478&
0.6652&
6500&
6465&
0.371&
\cite{2005AN....326...43G}\\
69&
0.3133885(5)&
0.130&
6300&
6215(55)&
9.346&
\cite{2012NewA...17..603E}\\
70&
0.38299992(8)&
0.81214(24)&
6240&
6237&
0.1896&
\cite{2019RAA....19...56H}\\
71&
0.30342695(3)&
0.5491&
4962&
5293(6)&
0.533(2)&
\cite{2013NewA...19...27Y}\\
72&
0.46698086(4)&
0.8162(9)&
6650(275)&
6578(271)&
0.190(6)&
\cite{2017PASJ...69...69Y}\\
73&
0.47067725(4)&
0.6733(17)&
5940(125)&
5910(124)&
0.455(20)&
\cite{2017PASJ...69...69Y}\\
74&
0.38999741(2)&
0.270(2)&
5746(33)&
5420(56)&
2.68(6)&
\cite{2013AJ....146..157C}\\
75&
0.2052(8)&
0.6987(15)&
3840&
3801(4)&
0.43(1)&
\cite{2015RAA....15.2237J}\\
        \hline
\end{tabular}
\end{center}
\label{tab9}
\end{table*}

\begin{table*}
\renewcommand\thetable{tab9}
\caption{Continued} 
\centering
\begin{center}
\footnotesize
\begin{tabular}{c c c c c c c c }
 \hline
 \hline
         System & $P$(day) & $l_{1}/l_{tot}(V)$ & $T_{1}(K)$ & $T_{2}(K)$ & $q$ & Reference\\
		\hline
76&
0.4350142(1)&
0.530(8)&
6255&
6150(9)&
0.896(3)&
\cite{2019RAA....19..174H}\\
77&
0.48068667(2)&
0.657&
6900&
6796(66)&
0.491&
\cite{2004AA...415..283D}\\
78&
0.4756119(2)&
0.797&
6095&
5946(12)&
0.226&
\cite{2006PASA...23..154D}\\
79&
0.2840975(3)&
0.436&
4850&
4524(10)&
1.640&
\cite{2006PASA...23..154D}\\
80&
0.30699175(1)&
0.3281(45)&
5568&
5281&
2.90(6)&
\cite{2017PASP..129l4202W}\\
81&
0.32741320(7)&
0.365(2)&
5326&
4995(11)&
2.300(1)&
\cite{2015NewA...38...50Z}\\
82&
0.22512639(3)&
0.4590(121)&
4680&
4523(21)&
1.5488(163)&
\cite{2018RAA....18...30Z}\\
83&
0.39634359(10)&
0.3146(6)&
6210&
6121(3)&
2.554(4)&
\cite{2011AJ....141...44L}\\
84&
0.485952&
0.8808(1)&
7091&
5414(10)&
0.430(4)&
\cite{2020PASA...37...31T}\\
85&
0.28980595(9)&
0.7169(16)&
5800(80)&
5976(8)&
0.305(8)&
\cite{2018RAA....18...78Y}\\
86&
0.28233519(8)&
0.622(6)&
5498(200)&
5497(209)&
0.548(19)&
\cite{2021AstL...47..402P}\\
87&
0.39972197(5)&
0.863(55)&
5760&
5875(159)&
0.082&
\cite{2021OAst...30...37P}\\
88&
0.250759986(28)&
0.743(1)&
5264&
5275(12)&
0.293(3)&
\cite{2021RAA....21..193G}\\
89&
0.28895094(43)&
0.3944(24)&
5197(12)&
4920&
2.281(34)&
\cite{2021RAA....21..122L}\\
90&
0.29715993(16)&
0.6558(60)&
5890&
5873(10)&
0.207(5)&
\cite{2011AJ....142...99C}\\
91&
0.5315423(4)&
0.8378(1)&
7080&
6771(3)&
0.192(1)&
\cite{2018JAVSO..46..133A}\\
92&
0.42439453&
0.5614(8)&
5310&
5906(8)&
0.427(9)&
\cite{2012NewA...17..347D}\\
93&
0.318194081(114)&
0.7818(6)&
5873&
5850(35)&
0.2074(1)&
\cite{2021NewA...8601571P}\\
94&
0.5932963(1)&
0.702&
7580&
7559(15)&
0.36&
\cite{2005NewA...10..163A}\\
95&
0.421523(2)&
0.782&
6500&
6180(10)&
0.277(24)&
\cite{2014AJ....148..126C}\\
96&
0.597423(2)&
0.818&
7000&
6550(20)&
0.250(5)&
\cite{2014AJ....148..126C}\\
97&
0.305371(2)&
0.606&
6200&
6400(10)&
0.534(6)&
\cite{2014AJ....148..126C}\\
98&
0.375307(3)&
0.751&
6250&
6280(10)&
0.286&
\cite{2014AJ....148..126C}\\
99&
0.2897446&
0.27(1)&
5074&
5600&
5.41(2)&
\cite{2016AJ....152..129C}\\
100&
0.3069522&
0.41(1)&
5275&
4975&
2.09(1)&
\cite{2016AJ....152..129C}\\
        \hline
\end{tabular}
\end{center}
\label{tab9}
\end{table*}

\begin{table*}
\renewcommand\thetable{tab9}
\caption{Continued} 
\centering
\begin{center}
\footnotesize
\begin{tabular}{c c c c c c c c }
 \hline
 \hline
         System & $P$(day) & $l_{1}/l_{tot}(V)$ & $T_{1}(K)$ & $T_{2}(K)$ & $q$ & Reference\\
		\hline
101&
0.3424570&
0.61(1)&
5582&
5308&
0.78(1)&
\cite{2016AJ....152..129C}\\
102&
0.3281916&
0.47(2)&
5546&
5088&
1.90(1)&
\cite{2016AJ....152..129C}\\
103&
0.3304319&
0.68(2)&
5383&
5175&
0.52(2)&
\cite{2016AJ....152..129C}\\
104&
0.3581278&
0.46(2)&
5750&
5370&
1.60(2)&
\cite{2016AJ....152..129C}\\
105&
0.40589680(8)&
0.605(3)&
4899(124)&
5109(139)&
0.475&
\cite{2019PASP..131h4202L}\\
106&
0.41745021(25)&
0.7023(3)&
6250(1)&
6379(1)&
0.2629(3)&
\cite{2018JAVSO..46...57S}\\
107&
0.40410880(3)&
0.3031(6)&
6250&
6176(5)&
2.69438(1162)&
\cite{2013AJ....146...79L}\\
108&
0.43035823(6)&
0.7394(8)&
6150&
6201(18)&
0.297&
\cite{2007AJ....133..357Q}\\
109&
0.38420990(8)&
0.6848(18)&
6700&
6720(42)&
0.41&
\cite{2008AJ....136.2493Q}\\
110&
0.50055604(8)&
0.6201(4)&
6400&
6316(32)&
0.610(3)&
\cite{2014AJ....148...79Q}\\
111&
0.37421962(7)&
0.5991(5)&
6104&
6003(5)&
0.693(6)&
\cite{2020RAA....20...50W}\\
112&
0.44974752(21)&
0.7140(9)&
5721(9)&
6100&
0.257(1)&
\cite{2004PASP..116..826Y}\\
113&
0.47606620(3)&
0.8211(9)&
6352&
6116(12)&
0.211(6)&
\cite{2015AJ....150...83Z}\\
114&
0.26267695(1)&
0.3316(8)&
5113(5)&
4973&
2.61(4)&
\cite{2020RAA....20...10Z}\\
115&
0.36045754(20)&
0.830(1)&
6300&
6354(7)&
0.149(1)&
\cite{2005AJ....129..979Z}\\
116&
0.344909289(3)&
0.372&
6150&
5885(50)&
2.278&
\cite{2005MNRAS.363.1272Y}\\
117&
0.3234561(4)&
0.6857(12)&
5648&
5873(7)&
0.3425(14)&
\cite{2016RMxAA..52..339M}\\
118&
0.29225601(9)&
0.335&
4830&
4581(5)&
2.84533(609)&
\cite{2012AJ....144..178L}\\
        \hline
        \hline
\end{tabular}
\end{center}
\label{tab9}
\end{table*}

\clearpage
Note: 1=UY UMa, 2=BF Pav, 3=V2240 Cyg, 4=DF CVn, 5=GSC 1042-2191, 6=AA UMa, 7=V789 Her, 8=VZ Psc, 9=ASAS J212234-4627.6, 10=ASAS J212319-4622.4, 11=V608 Cas, 12=QX And, 13=ASAS J174406+2446.8, 14=V1848 Ori, 15=QQ Boo, 16=1SWASP J064501.21+342154.9, 17=NSVS 2701634, 18=V1370 Tau, 19=V1007 Cas, 20=GW Cnc, 21=2MASS 02272637+1156494, 22=NX Cam, 23=NSVS 2643686, 24=V584 Cam, 25=GR Vir, 26=FG Hydrae, 27=UX Eri, 28=RT LMi, 29=NSVS 7245866, 30=NR Cam, 31=V776 Cas, 32=PY Vir, 33=RZ Tau, 34=EH CVn, 35=EF CVn, 36=ASAS J083241+2332.4, 37=1SWASP J155822.10-025604.8, 38=1SWASP J212808.86+151622.0, 39=UCAC4 436-062932, 40=NSVS 1557555, 41=GSC 2723-2376, 42=MU Cnc, 43=TYC 1597-2327-1, 44=GSC 4946-0765, 45=V596 Peg, 46=V658 Lyr, 47=V700 Cyg, 48=FZ Ori, 49=LP UMa, 50=FP Lyn, 51=FV CVn, 52=V354 UMa, 53=OQ Cam, 54=EQ Tau, 55=V737 Per, 56=V336 TrA, 57=HN Psc, 58=V685 Peg, 59=1SWASP J080150.03+471433.8, 60=AB And, 61=GZ And, 62=AO Cam, 63=DK Cyg, 64=GSC 3581-1856, 65=J015829.5+260333, 66=J030505.1+293443, 67=KW Psc, 68=LO And, 69=V1191 Cyg, 70=V1853 Ori, 71=CE Leo, 72=V532 Mon, 73=GU Ori, 74=FI Boo, 75=SDSS J012119.10-001949.9, 76=AV Pup, 77=NN Vir, 78=AQ Psc, 79=XY Leo, 80=V2284 Cyg, 81=GK Aqr, 82=1SWASP J140533.33+114639.1, 83=V396 Mon, 84=TT Cet, 85=PS Vir, 86=BQ Ari, 87=V870 Ara, 88=V811 Cep, 89=V0842 Cep, 90=TZ Boo, 91=GW Boo, 92=BE Cep, 93=BO Ari, 94=V351 Peg, 95=AK Her, 96=HI Dra, 97=V1128 Tau, 98=V2612 Oph, 99=EP Cep, 100=EQ Cep, 101=ES Cep, 102=V369 Cep, 103=V370 Cep, 104=V782 Cep, 105=BH Cas, 106=V573 Peg, 107=EP And, 108=AP Leo, 109=BI CVn, 110=AL Cas, 111=V680 Per, 112=EZ Hydrae, 113=MQ UMa, 114=V1197 Her, 115=AH Cnc, 116=V781 Tau, 117=ROTSE1 J164341.65+251748.1, 118=GSC 03526-01995.

\clearpage
 
\bibliographystyle{aasjournal}
\bibliography{new.ms}

\end{document}